\documentclass[%
 reprint,
 amsmath,amssymb,
 aps,
]{revtex4-2}

\usepackage{graphicx}
\usepackage{dcolumn}
\usepackage{bm}
\usepackage{bm}
\usepackage{subcaption} 
\usepackage[justification=raggedright,singlelinecheck=false]{caption}

\begin{document}

\preprint{APS/123-QED}

\title{Systematic Study on the $\alpha$-particle preformation factor in the theory of $\alpha$-decay based on the Tabular Prior-data Fitted Network (TabPFN)}

\author{Panpan Qi}
\affiliation{College of Physics and Technology, Kunming University, Kunming 650214, China.}
\author{Xuanpeng Xiao}%
\affiliation{College of Physics and Technology, Kunming University, Kunming 650214, China.}%

\author{Gongming Yu}%
 \email{ygmanan@kmu.edu.cn}
\affiliation{College of Physics and Technology, Kunming University, Kunming 650214, China.}%

\author{Haitao Yang}%
 \email{yanghaitao205@163.com}
\affiliation{ College of Science, Zhaotong University, Zhaotong 657000, China.}%

\author{Qiang Hu}%
 \email{qianghu@impcas.ac.cn}
\affiliation{Institute of Modern Physics, Chinese Academy of Sciences, Lanzhou 730000, China.}%

\begin{abstract}
A hybrid approach combining the Tabular Prior-data Fitted Network (TabPFN) with the Coulomb and Proximity Potential Model (CPPM) is developed to investigate $\alpha$-particle preformation factors $P_{\alpha}$ and their impact on $\alpha$-decay half-lives. The TabPFN model, trained on 498 nuclei, accurately learns the relationship between nuclear structure properties and $P_{\alpha}$, achieving a root mean square deviation of $\sigma_{\mathrm{rms}} = 0.211$. The predicted factors reveal clear odd-even staggering and shell closure effects, and exhibit linear correlations with both $Q_{\alpha}^{-1/2}$ and the fragmentation potential $V_{\mathrm{frag}}$. When incorporated into CPPM calculations, the machine-learning-based $P_{\alpha}$ values significantly improve half-life predictions. Similar improvements are also obtained when deformation effects are included in the potential barrier description. The capability of the model is further demonstrated through predictions for superheavy nuclei ($Z = 117$--120), suggesting $N = 184$ as a potential neutron magic number.

\end{abstract}

\maketitle


\section{\label{sec:level1}Introduction}

The phenomenon of $\alpha$-decay was first discovered by Rutherford in 1899. In 1928, Gamow and Condon theoretically established that $\alpha$-decay originates from the quantum tunneling effect \cite{gamow1928quantentheorie,gurney1928wave}. Since then, $\alpha$-decay has been recognized as a pivotal tool for investigating the structure of unstable atomic nuclei, offering critical insights into their internal dynamics.

Consequently, the development of a robust theoretical framework for $\alpha$-decay is of paramount importance. Advances in $\alpha$-decay theory have primarily addressed aspects including potential barrier profiles, decay energies \cite{jin2008alpha,ni2012binding}, nuclear medium effects \cite{deng2019significant}, nuclear deformation \cite{guo2015nuclear}, and the $\alpha$-particle preformation factor \cite{zhang2009preformation,2016Systematics}. Furthermore, nuclear deformation may influence the barrier penetration probability in $\alpha$ decay, since the quadrupole deformation of the Coulomb barrier can modify both the height and the width of the potential barrier \cite{delion2010theory}. Research on $\alpha$-decay theory has yielded numerous achievements, such as empirical formulas \cite{viola1966nuclear,ren2012new,royer2000alpha,brown1992simple,qi2009universal}, macro-microscopic models, and microscopic approaches \cite{royer2000alpha,royer1985static,royer2010analytic,joseph2018theoretical,sun2016systematic,deng2017systematic,dong2010alpha,xu2006global}.

In recent years, machine learning (ML) algorithms have emerged as prominent alternative tools for investigating and predicting complex physical data, owing to their strong performance and increasingly vital role in this domain \cite{lecun2015deep,schmidhuber2015deep}. In nuclear physics, ML techniques have been successfully applied to predict nuclear masses \cite{carnini2020trees,wu2021nuclear,wu2020predicting}, binding energies \cite{murarka2022neutron}, charge radii \cite{wu2020calculation}, $\beta$-decay lifetimes \cite{clark2006application}, and $\alpha$-decay rates \cite{saxena2021modified}. In studies of heavy nuclei and density functional theory (DFT), ML facilitates the calibration of energy density functional (EDF) parameters \cite{bollapragada2020optimization,goriely2014uncertainties} and enables energy predictions for over a thousand nuclides with low computational cost \cite{lasseri2020taming}. In investigations of nuclear properties, ML is employed to assess model-related uncertainties in multi-model frameworks \cite{connell2021does}; in nuclear reaction and neutron star studies, research focuses on the equation of state (EoS) of nuclear matter \cite{margueron2018equationa,margueron2018equationb}.

Deep learning techniques have been employed in studies of $\alpha$-decay half-lives \cite{you2024calculating,shree2025alpha,jalili2024decay,jin2023bayesian}. In Ref. \cite{you2024calculating}, an artificial neural network (ANN) was utilized to compute $\alpha$-decay half-lives; incorporating the angular momentum carried by $\alpha$ particles and the quadrupole deformation parameters of parent nuclei reduced the root-mean-square deviation (RMSD) between predictions and experimental data from 0.581 to 0.334. In Ref. \cite{shree2025alpha}, the XGBoost algorithm was adopted to predict $\alpha$-decay half-lives of superheavy nuclei, demonstrating the smallest mean square deviation compared to empirical formulas and experimental data. In Ref. \cite{jalili2024decay}, the radial basis function (RBF) kernel support vector machine (SVM) was applied to predict nuclear $\alpha$-decay half-lives; analysis of 2232 data points showed excellent performance, with root-mean-square errors (RMSE) of 0.819 and 0.352 on two datasets, comparable to other ML methods. In Ref. \cite{jin2023bayesian}, a Bayesian neural network (BNN) combined with a complex $\alpha$-decay model was proposed, leading to improved half-life predictions through modifications to the preformation factor.

Within the framework of $\alpha$-decay theory, the $\alpha$-particle preformation factor characterizes the probability of $\alpha$-cluster formation on the surface of the parent nucleus. Typically, $P_{\alpha}$ is treated as an empirical constant, based on its smooth variation in open-shell regions \cite{hodgson2003cluster}; however, this assumption may introduce inaccuracies in reproducing half-lives for nuclides near closed-shell regions. In closed-shell nuclides (e.g., those with proton number $Z = 82$ and neutron number $N = 126$), $P_{\alpha}$ exhibits significant variations \cite{andreyev2013signatures,deng2015realistic}, potentially leading to systematic deviations between theoretical and experimental $\alpha$-decay half-lives. Recent work by Zhang and Deng \cite{deng2021correlation} demonstrated progress by extracting preformation factors via a semi-empirical formula, enhancing consistency between predictions and observations. The discrepancy between theoretical and experimental half-lives primarily stems from the treatment of $P_{\alpha}$, and given ML's unique capability to address discrepancies between datasets and extract physical insights, its potential for investigating $\alpha$-particle preformation factors warrants further exploration.

This work investigates the capability of the Tabular Prior-data Fitted Network (TabPFN) to extract $\alpha$-particle preformation factors, aiming to enhance predictions of $\alpha$-decay half-lives. Section II outlines the theoretical framework, describing the extraction of experimental preformation factors via the Coulomb and Proximity Potential Model (CPPM) and the TabPFN methodology. In Section III, we demonstrate that the TabPFN model effectively learns complex relationships between preformation factors and nuclear structure properties, capturing key features such as odd-even staggering and shell closure effects, and revealing a linear correlation with $Q_\alpha^{-1/2}$. The model's extrapolation capability is validated against 41 newly evaluated nuclei, significantly improving the accuracy of predicted half-lives. Finally, the model is employed to predict preformation factors and half-lives for superheavy nuclei ($Z = 117 - 120$), suggesting $N=184$ as a potential neutron magic number. A summary is provided in Section IV.

\section{\label{sec:level2}THEORETICAL FRAMEWORK}

\subsection{\label{sec:level2}The extraction of experimental preformation factors $P_{\alpha}$ exp under the framework of CPPM}

The $\alpha$-decay half-life of atomic nuclei is given by \cite{qi2023systematic}:
\begin{eqnarray}\label{y1}
T_{1/2} = \frac{\ln 2}{\lambda},
\end{eqnarray}
where $\lambda$ denotes the decay constant, expressed as
\begin{eqnarray}\label{y2}
\lambda = P_{\alpha} \nu P.
\end{eqnarray}
Here, $P_{\alpha}$, $\nu$, and $P$ represent the $\alpha$-particle preformation factor, assault frequency, and barrier penetration probability, respectively.

This study begins by examining the relationship between the $\alpha$-decay constant $\lambda$ and the half-life $T_{1/2}$. Combining Eqs. (1) and (2) with the experimentally measured half-life $T_{1/2}^{\text{exp}}$ yields
\begin{eqnarray}\label{y3}
\lambda_{\text{exp}} = \frac{\ln 2}{T_{1/2}^{\text{exp}}} = P_\alpha^{\text{exp}} \nu P ,
\end{eqnarray}
\begin{eqnarray}\label{y4}
\lambda_{\text{cal}} = \frac{\ln 2}{T_{1/2}^{\text{cal}}} = P_\alpha^{\text{cal}} \nu P .
\end{eqnarray}
Within the unified decay theory framework, setting the theoretical preformation factor $P_\alpha^{\text{cal}}$ to unity allows extraction of the experimental preformation factor as:
\begin{eqnarray}\label{y5}
P_{\alpha}^{\text{exp}} = \frac{T_{\alpha}^{\text{cal}}}{T_{\alpha}^{\text{exp}}}.
\end{eqnarray}

The Coulomb and Proximity Potential Model (CPPM) is employed as the unified decay theory framework, owing to its accurate description of nuclear fusion, fission, and particularly $\alpha$ decay. The assault frequency $\nu$ is set to $1.0 \times 10^{22} , \text{s}^{-1}$ \cite{poenaru2011single}. The penetration probability $P$ is computed using the semiclassical Wentzel-Kramers-Brillouin (WKB) approximation as
\begin{eqnarray}\label{y6}
P = -\frac{2}{\hbar}\int_{R_{\text{in}}}^{R_{\text{out}}} \sqrt{2\mu \left( V(r) - Q_\alpha \right)} , dr.
\end{eqnarray}
Here, $\hbar$ denotes the reduced Planck constant, $\mu = \frac{m_d m_\alpha}{m_d + m_\alpha}$ is the reduced mass of the $\alpha$-daughter system, with $m_d$ and $m_\alpha$ representing the masses of the daughter nucleus and $\alpha$ particle, respectively, and $Q_\alpha$ is the $\alpha$-decay energy.

The total interaction potential between the emitted $\alpha$ particle and daughter nucleus consists of nuclear, Coulomb, and centrifugal terms:
\begin{eqnarray}\label{y7}
V(r) = V_N(r) + V_C(r) + V_\ell(r).
\end{eqnarray}
Here, $r$ is the separation between the centers of the daughter nucleus and $\alpha$ particle. The nuclear potential is described using the proximity potential formalism proposed by Blocki \textit{et al.} \cite{blocki1977proximity}:
\begin{eqnarray}\label{y8}
V_N(r) = 4\pi \gamma b\overline{R}\Phi(\xi).
\end{eqnarray}
where $\gamma = \gamma_0 [1 - k_s (\frac{N_{p} - Z_{p}}{N_{p} + Z_{p}})^2]$ is the surface energy coefficient, with $\gamma_0 = 0.9517$ MeV/fm$^2$ and $k_s = 1.7826$ representing the surface energy constant and surface asymmetry coefficient, respectively. Here, $N_{p}$ and $Z_{p}$ denote the neutron and proton numbers of the parent nucleus, respectively; $b = 1$ is the nuclear surface diffuseness, and $\overline{R}$ is the mean curvature radius, expressed as
\begin{eqnarray}\label{y9}
\overline{R} = \frac{C_{\alpha} C_{d}}{C_{\alpha} + C_{d}}.
\end{eqnarray}
The quantity $C_i$ ($i = \alpha, d$) is given by $C_i = R_i \left[ 1 - \left( \frac{b}{R_i} \right)^2 \right]$, where $C_{\alpha}$ and $C_{d}$ are the matter radii of the $\alpha$ particle and daughter nucleus, respectively. The effective sharp radius $R_i$ ($i = \alpha, d$) is \cite{royer2000alpha}: $R_i = 1.28A_i^{1/3} - 0.76 + 0.8A_i^{-1/3}$, where $A_i$ is the mass number of the $\alpha$ particle ($i = \alpha$) or daughter nucleus ($i = d$).

The universal function $\Phi(\xi)$ is defined as
\begin{align}
\phi(\xi) = \begin{cases}
\frac{1}{2}(\xi - 2.54)^2 - 0.0852(\xi - 2.54)^3, & \xi < 1.2511, \\
-3.437 \exp\left(-\frac{\xi}{0.75}\right), & \xi \geq 1.2511.
\end{cases}
\end{align}
with $\xi = (r - C_{\alpha} - C_{d})/b$ representing the separation between the near surfaces of the daughter nucleus and $\alpha$ particle.

The Coulomb potential $V_c(r)$ is that of a uniformly charged sphere of radius $R$:
\begin{eqnarray}\label{y11}
V_c(r) =
\begin{cases}
\frac{Z_\alpha Z_d e^2}{2R} \left[ 3 - \left( \frac{r}{R} \right)^2 \right], & r \leq R, \\
\frac{Z_\alpha Z_d e^2}{r}, & r > R.
\end{cases}
\end{eqnarray}
Here, $R = R_d + R_\alpha$ is the sum of the radii of the daughter nucleus and $\alpha$ particle, and $Z_d$ and $Z_\alpha$ are the proton numbers of the daughter nucleus and $\alpha$ particle, respectively.

The centrifugal potential $V_l(r)$ employs the Langer-modified form, where $l(l+1) \to \left( l + \frac{1}{2} \right)^2$, essential for one-dimensional problems \cite{morehead1995asymptotics}:
\begin{equation}\label{y11}
V_l(r) = \frac{\hbar^2 \left( l + \frac{1}{2} \right)^2}{2 \mu r^2},
\end{equation}
where $l$ is the orbital angular momentum carried by the emitted $\alpha$ particle. The minimum $l$ is determined by conservation of parity and angular momentum.

For the integral bounds in Eq. (6), the outer turning point radius is $R_{\text{out}} = \frac{Z_d Z_\alpha e^2}{2Q_\alpha} + \sqrt{\left( \frac{Z_d Z_\alpha e^2}{2Q_\alpha} \right)^2 + \frac{\hbar^2 l(l+1)}{2\mu Q_\alpha}}$, and the inner turning point is $R_{\text{in}} = R_d + R_\alpha$, corresponding to the separation configuration \cite{dong2009cluster}.

The above expressions are obtained under the assumption of spherical symmetry. However, many nuclei involved in $\alpha$ decay exhibit quadrupole deformation, which may modify the Coulomb barrier and consequently influence the barrier penetration probability. In order to take this effect into account, the Coulomb interaction is extended to the case of axially symmetric deformed nuclei. Following the treatment in Ref.~\cite{dahmardeh2017diffuseness}, the deformed Coulomb potential can be written as

\begin{equation}
V_C(r,\theta)=\frac{Z_\alpha Z_d e^2}{4\pi\varepsilon_0 r}
\left(1-e^{-\phi(\theta)r-0.5[\phi(\theta)r]^2-0.35[\phi(\theta)r]^3}\right),
\end{equation}
The deformation parameter $\phi(\theta)$ is related to the deformed nuclear radius through
\begin{equation}
\phi(\theta)R(\theta)=\frac{3}{2},
\end{equation}
where the deformed nuclear radius is expressed as

\begin{equation}
R(\theta)=r_0A_d^{1/3}\left[1+\beta_2Y_{20}(\theta)+\beta_4Y_{40}(\theta)\right].
\end{equation}
The nuclear potential, transmission coefficient, and detailed 
calculation procedure for the deformed barrier penetration follow the approach described in Ref.~\cite{dahmardeh2017diffuseness}.

\subsection{\label{sec:citeref}Tabular Prior-data Fitted Network method}
The Tabular Foundation Model (TabPFN) is employed—a Transformer-based architecture specifically designed for small-to-medium-sized tabular datasets (containing up to 10,000 samples and 500 features). Consider a single tabular dataset $\mathcal{D} = {(x_i, y_i)}_{i=1}^N$ comprising $N$ training samples (table rows). Each instance $x_i$ is characterized by $d$ features (columns), where $d$ may vary across datasets. The label $y$ belongs to $[\mathcal{C}] = {1, \ldots, C}$ for classification tasks or is a numerical value for regression tasks. All attributes are assumed numerical (continuous); categorical (discrete) attributes, if present, are preprocessed using ordinal or one-hot encoding. The objective is to learn a mapping $f$ from instances to their corresponding labels. Specifically, given an unseen instance $x^* \in \mathbb{R}^d$ sampled from the same distribution as $\mathcal{D}$, the learned mapping $f$ predicts its label as $\hat{y}^{*} = f(x^* \mid \mathcal{D})$. A smaller discrepancy between $\hat{y}^{*}$ and the true label $y^{*}$ indicates stronger generalization capability \cite{ye2025closer}.

The complete input to TabPFN —comprising both the training set and the test instance—forms a tensor of shape $(N + 1) \times (d + 1) \times k$. The model alternately applies two types of self-attention: across samples (among the $N{+}1$ instances) and across attributes (among the $d{+}1$ dimensions), enabling in-context learning along both axes \cite{hollmann2025accurate,ye2025closer}. Finally, the output token corresponding to the test instance's dummy label $\tilde{y}^*$ is extracted and mapped either to a 10-class logit (classification) or a single-value logit (regression).

The weights of TabPFN  are pretrained on diverse synthetic datasets generated via structural causal models (SCMs), with checkpoints selected according to performance on real-world datasets. Further details regarding feature preprocessing, acceleration techniques, and post-hoc ensembling are provided in Ref. \cite{hollmann2025accurate}.

In practical applications, 498 nuclear features are input to the TabPFN model, including ten parameters: mass number $A$, proton number $Z$, neutron number $N$, $\alpha$-decay energy $Q_\alpha$, minimum angular momentum $l$, quadrupole deformation $\beta_2$ of the parent nucleus \cite{moller1993nuclear,denisov2025alpha}, parity of $Z$ and $N$ ($Z_p$, $N_p$), and pairing effect $\delta$ \cite{kirson2008mutual}. Leveraging the in-context learning (ICL) mechanism, the model processes the entire dataset in a single forward pass. TabPFN is pretrained on millions of synthetic datasets generated via structural causal models (SCMs). After pretraining, TabPFN can be directly applied to new real-world datasets without further training.

\subsection{Effect of Ten Nuclear Physics Descriptors on Alpha Decay Preformation Factor with TabPFN Model}

\begin{table*}[]
\caption{The $\sigma_{RMS}$ for the training set (399 nuclei) and testing set (99 nuclei) with different TabPFN models.}
\centering
\begin{tabular}{ccccc}
\hline
\multicolumn{2}{c}{Inputs}                              & $\sigma_{RMS}$\text{(Testing set)} & $\sigma_{RMS}$\text{(Training set)} & $\sigma_{RMS}$\text{(Entire set)} \\ \hline
TabPFN1  & $A$, $Z$, $N$, $Q_{\alpha}$ & 0.511                       & 0.379                       & 0.409                     \\
TabPFN2  & $A$, $Z$, $N$, $\delta$, $Q_{\alpha}$                 & 0.361                       & 0.236                        & 0.266                      \\
TabPFN3  & $A$, $Z$, $N$, $Z_{p}$, $N_{p}$, $Q_{\alpha}$                       & 0.358                       & 0.218                        & 0.252                      \\
TabPFN4  & $A$, $Z$, $N$, $Z_{p}$, $N_{p}$, $\delta$, $Q_{\alpha}$               & 0.356                       & 0.208                        & 0.245                        \\
TabPFN5  & $A$, $Z$, $N$, $l$, $Q_{\alpha}$                     & 0.389                       & 0.262                        & 0.292                      \\
TabPFN6  & $A$, $Z$, $N$, $l$, $\delta$, $Q_{\alpha}$              & 0.328                       & 0.212                        & 0.24                      \\
TabPFN7  & $A$, $Z$, $N$, $l$,  $Z_{p}$, $N_{p}$, $Q_{\alpha}$                 & 0.375                       & 0.164                        & 0.247                      \\
TabPFN8  & $A$, $Z$, $N$, $l$, $Z_{p}$, $N_{p}$, $\delta$, $Q_{\alpha}$           & 0.304                        & 0.195                        & 0.222                      \\
TabPFN9  & $A$, $Z$, $N$, $l$, $\beta_2$, $Q_{\alpha}$                  & 0.423                       & 0.265                        & 0.285                       \\
TabPFN10 & $A$, $Z$, $N$, $l$, $\delta$, $\beta_2$, $Q_{\alpha}$         & 0.3                       & 0.206                        & 0.228                      \\
TabPFN11 & $A$, $Z$, $N$, $l$,  $Z_{p}$, $N_{p}$, $\beta_2$, $Q_{\alpha}$            & 0.313                       & 0.184                        & 0.216                      \\
TabPFN12 & $A$, $Z$, $N$, $l$, $Z_{p}$, $N_{p}$, $\delta$, $\beta_2$, $Q_{\alpha}$        & 0.247                       & 0.197                        & 0.211                      \\ \hline
\end{tabular}
\end{table*}

In nuclear physics research, machine learning methods can be roughly categorized into direct learning and indirect learning. The core objective of direct learning is to obtain experimental values of physical targets, thereby verifying the recognition performance of machine learning models for physical targets; indirect learning optimizes theoretical calculation models by analyzing the deviations between experimental values and theoretical values. In this study, the CPPM (with $P_\alpha=1$) was used to calculate $\alpha$-decay half-lives, and experimental preformation factors were extracted from 498 nuclides (including 172 even-even nuclei, 261 odd-A nuclei, and 65 odd-odd nuclei) according to Equation (5). The $\alpha$-decay half-lives and spin parameters used in the experiment were derived from the NUBASE2016 evaluation of nuclear properties \cite{audi2017nubase2016}, while the $\alpha$-decay energies were taken from the AME2016 atomic mass evaluation \cite{wang2021ame,kondev2017ame2016}. This study adopted an indirect learning strategy based on the Tabular Prior-data Fitted Network (TabPFN): preformation factors were first obtained indirectly via CPPM, followed by learning using TabPFN. To evaluate the consistency between the $\alpha$-particle preformation factors derived from TabPFN and those from experimental data, the root-mean-square deviation ($\sigma$rms) was employed as the metric to quantify model performance, defined as:

\begin{eqnarray}\label{y8}
\sigma_{\text{rms}} = \sqrt{\frac{1}{n} \sum_{i=1}^{n} \left( \log_{10} p_{{\alpha}i}^{\text{TabPFN}} - \log_{10} p_{{\alpha}i}^{\text{exp}} \right)^2}.
\end{eqnarray}

Here, $\log_{10} p_{{\alpha}i}^{\text{TabPFN}}$ denotes the logarithmic value of the preformation factor obtained via TabPFN learning, and $\log_{10} p_{{\alpha}i}^{\text{exp}}$ represents the logarithmic value of the experimental preformation factor extracted from the CPPM model, with n=498.

Table I presents the root-mean-square ($\sigma_{\mathrm{RMS}}$) deviations between the $\alpha$-particle preformation factors predicted by the TabPFN model and those extracted from experimental data across different datasets. Introducing the pairing effect $\delta$ in the TabPFN2 model reduces the $\sigma_{\mathrm{RMS}}$ value from 0.409 (TabPFN1) to 0.266, corresponding to a decrease of 34.9\%. Similarly, incorporating proton parity $Z_{\mathrm{P}}$ and neutron parity $N_{\mathrm{P}}$ in the TabPFN3 model leads to a $\sigma_{\mathrm{RMS}}$ value of 0.252, a reduction of 39.3\% compared to the TabPFN1 baseline. When pairing, proton parity, and neutron parity are simultaneously included in TabPFN4, the overall $\sigma_{\mathrm{RMS}}$ drops to 0.245, marking a 40\% improvement. These results demonstrate that including $\delta$, $Z_{\mathrm{P}}$, and $N_{\mathrm{P}}$ as input features significantly enhances the predictive performance of the TabPFN model.

Further physical features related to angular momentum and nuclear deformation were systematically investigated. Introducing the angular momentum parameter $l$ in the TabPFN5 model reduces the $\sigma_{\mathrm{RMS}}$ from 0.511 to 0.389 on the test set and from 0.379 to 0.262 on the training set, resulting in an overall reduction of 28.6\% relative to TabPFN1. This confirms that angular momentum effects substantially improve prediction accuracy. Subsequent models (TabPFN6, TabPFN7, and TabPFN8) also incorporate angular momentum, yielding further improvements: overall $\sigma_{\mathrm{RMS}}$ values decrease from 0.266 to 0.24, 0.252 to 0.247, and 0.245 to 0.222, respectively.

By additionally including the quadrupole deformation parameter $\beta_2$ of the parent nucleus in TabPFN9 - TabPFN11 models, the $\sigma_{\mathrm{RMS}}$ values are reduced from 0.292 to 0.285, 0.24 to 0.228, and 0.247 to 0.216, respectively. These results indicate that structural deformation plays a significant role in $\alpha$-preformation. The TabPFN10 model achieves $\sigma_{\mathrm{RMS}} = 0.228$, demonstrating excellent agreement with experimental preformation factors. The inclusion of $\beta_2$ also mitigates uncertainties related to magic and submagic numbers and provides insights into nuclear shell structure, offering a predictive approach for identifying shell closures in superheavy regions.

The full TabPFN12 model, incorporating all nine physical features, achieves the best performance with $\sigma_{\mathrm{RMS}} = 0.211$ across the entire dataset, confirming its capability to accurately reproduce experimental $\alpha$-preformation factors.

To further examine the role of nuclear deformation, the $\alpha$-particle preformation factors were also extracted using a deformed potential that combines the Coulomb interaction and the Woods–Saxon nuclear potential following the approach of Ref.~\cite{dahmardeh2017diffuseness}. The resulting preformation factors were subsequently predicted using the same TabPFN12 model. In this case, the obtained root-mean-square deviation is $\sigma_{\mathrm{RMS}}=0.232$, which remains close to the value obtained with the CPPM-based extraction. This agreement indicates that the TabPFN framework provides stable predictions for the $\alpha$-preformation factors even when different theoretical potentials are adopted.

\subsection{Preformation factor and nuclear structure information}

\begin{figure*}
    \centering
    \includegraphics[width=1\linewidth]{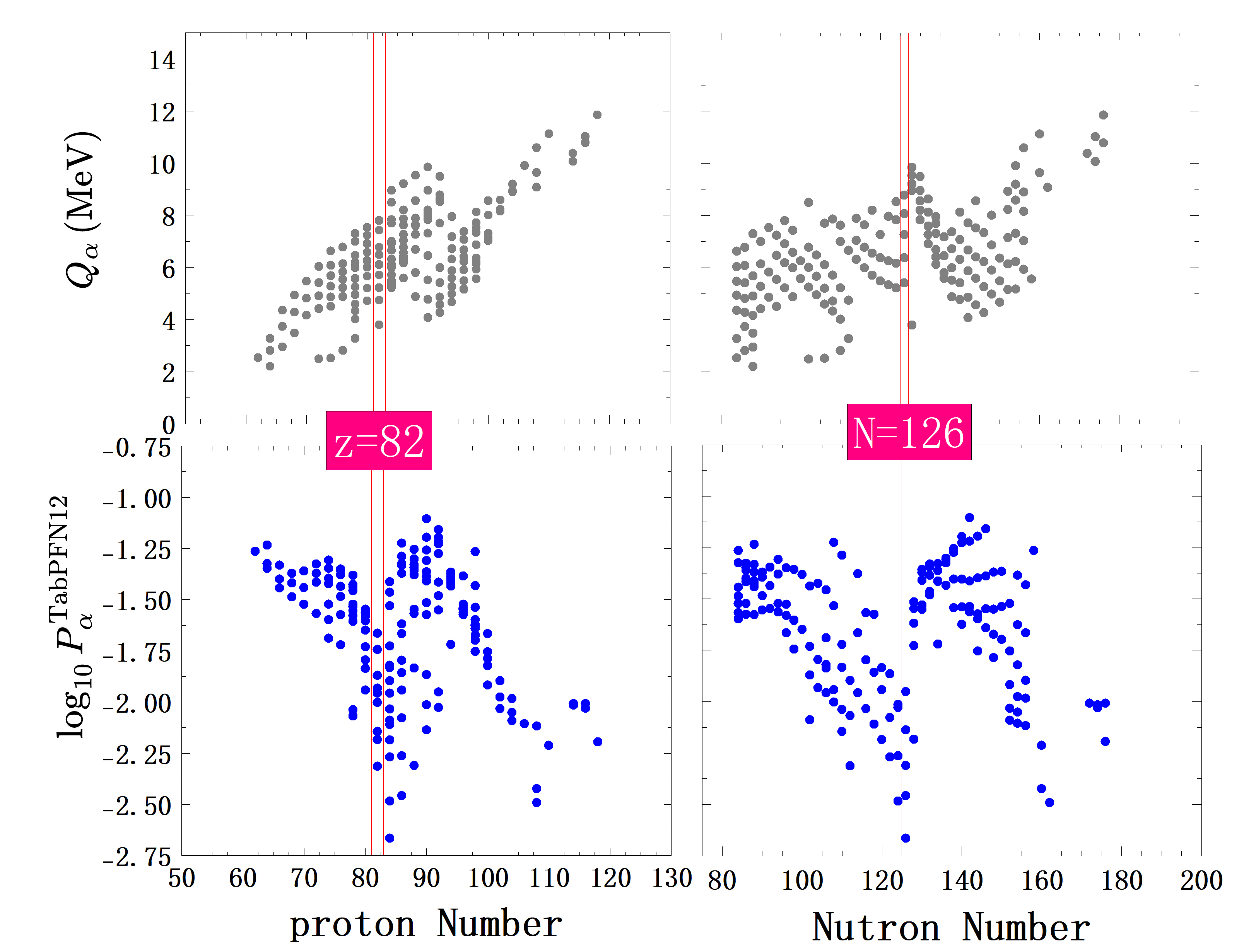}
    \caption{Upper panels: Experimental $\alpha$-decay energies $Q_\alpha$(in Mev)for even - even nuclei, plotted against proton number Z (left) and neutron number N (right). Lower panels: Logarithmic values of the $\alpha$-particle preformation factors $\log_{10} P_{\alpha}^{\mathrm{TabPFN12}}$ as functions of Z (left) and N (right).}
    \label{fig:1}
\end{figure*}

The TabPFN12 model, which exhibits optimal predictive performance, is employed to estimate the $\alpha$-particle preformation factors for 498 nuclei. The logarithmic preformation factors derived from the TabPFN12 method, along with the corresponding $\alpha$-decay energy $Q_\alpha$ (in MeV)—a key input to the network—are presented in Fig.~\ref{fig:1} as functions of the proton number $Z$ and neutron number $N$. For $Z < 82$ and $N < 126$, $Q_\alpha$ increases gradually with proton and neutron numbers, whereas $\log_{10} P_{\alpha}^{\text{TabPFN12}}$ decreases. Beyond these shell closures, both $Q_\alpha$ and $\log_{10} P_{\alpha}^{\text{TabPFN12}}$ exhibit significant variations. Specifically, beyond the $Z=82$ shell closure, $\log_{10} P_{\alpha}^{\text{TabPFN12}}$ initially rises before decreasing as $Z$ approaches the next proton magic number. A similar trend is observed for $N > 126$. Evidently, the trends of $\log_{10} P_{\alpha}^{\text{TabPFN12}}$ differ markedly across the $Z=82$ and $N=126$ closed shells, indicating that the preformation factor sensitively reflects nuclear shell effects. Thus, the preformation factor constitutes a reliable indicator for probing the stability of superheavy nuclei. Consistent with Ref. \cite{deng2021correlation}, the trend of $\log_{10} P_{\alpha}^{\text{TabPFN12}}$ confirms that the shell effect at $^{208}\text{Pb}$ ($Z=82$, $N=126$) is stronger than that at $^{132}\text{Sn}$ ($Z=50$, $N=82$). Furthermore, the highly regular and symmetric trends exhibited by the logarithmic preformation factor and the decay energy strongly support the empirical formula proposed in Ref. \cite{deng2021correlation}.

\begin{figure*}[htbp]
  \centering
  
  \begin{subfigure}{0.32\textwidth}
    \includegraphics[width=\linewidth]{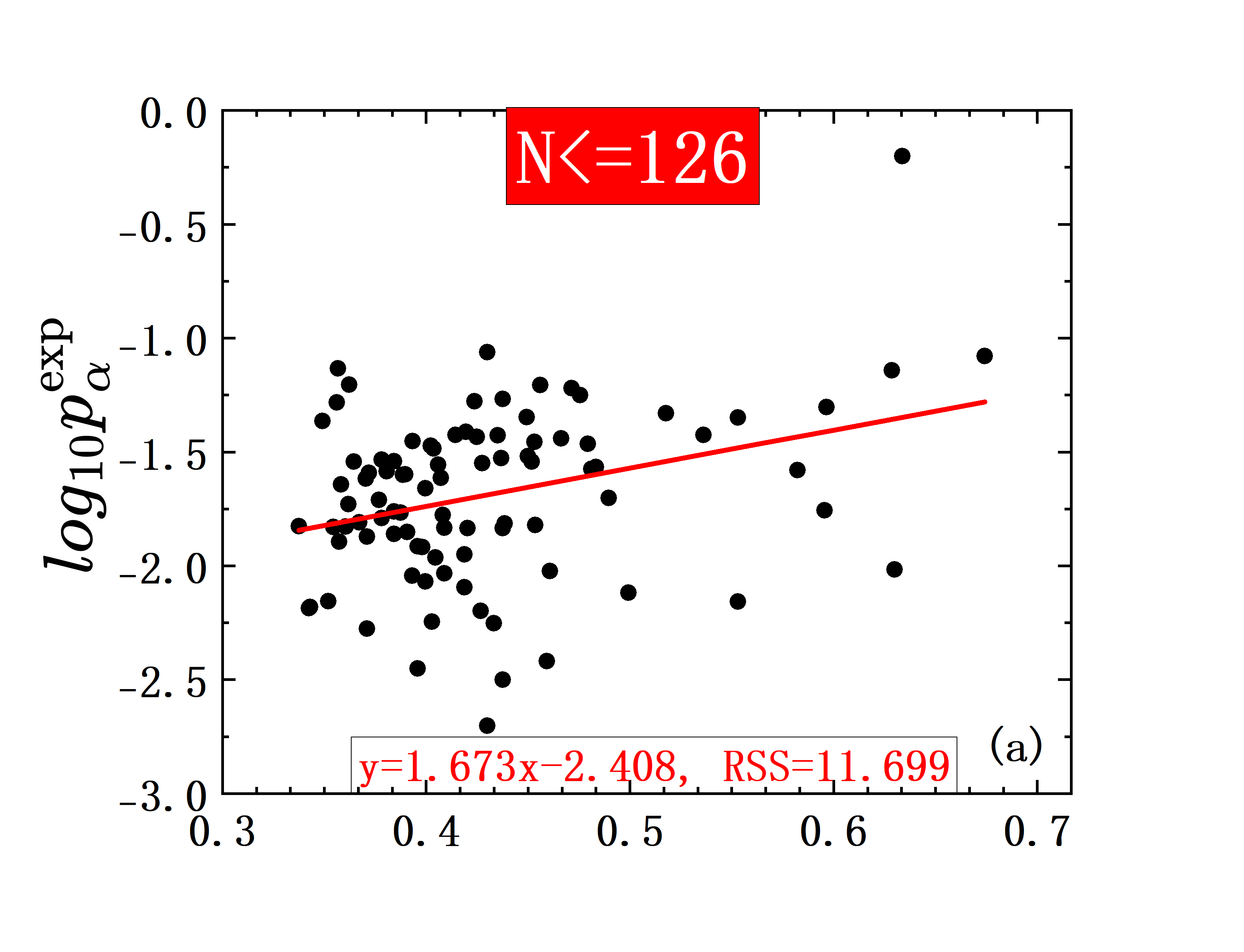}
  \end{subfigure}
  \hfill
  \begin{subfigure}{0.32\textwidth}
    \includegraphics[width=\linewidth]{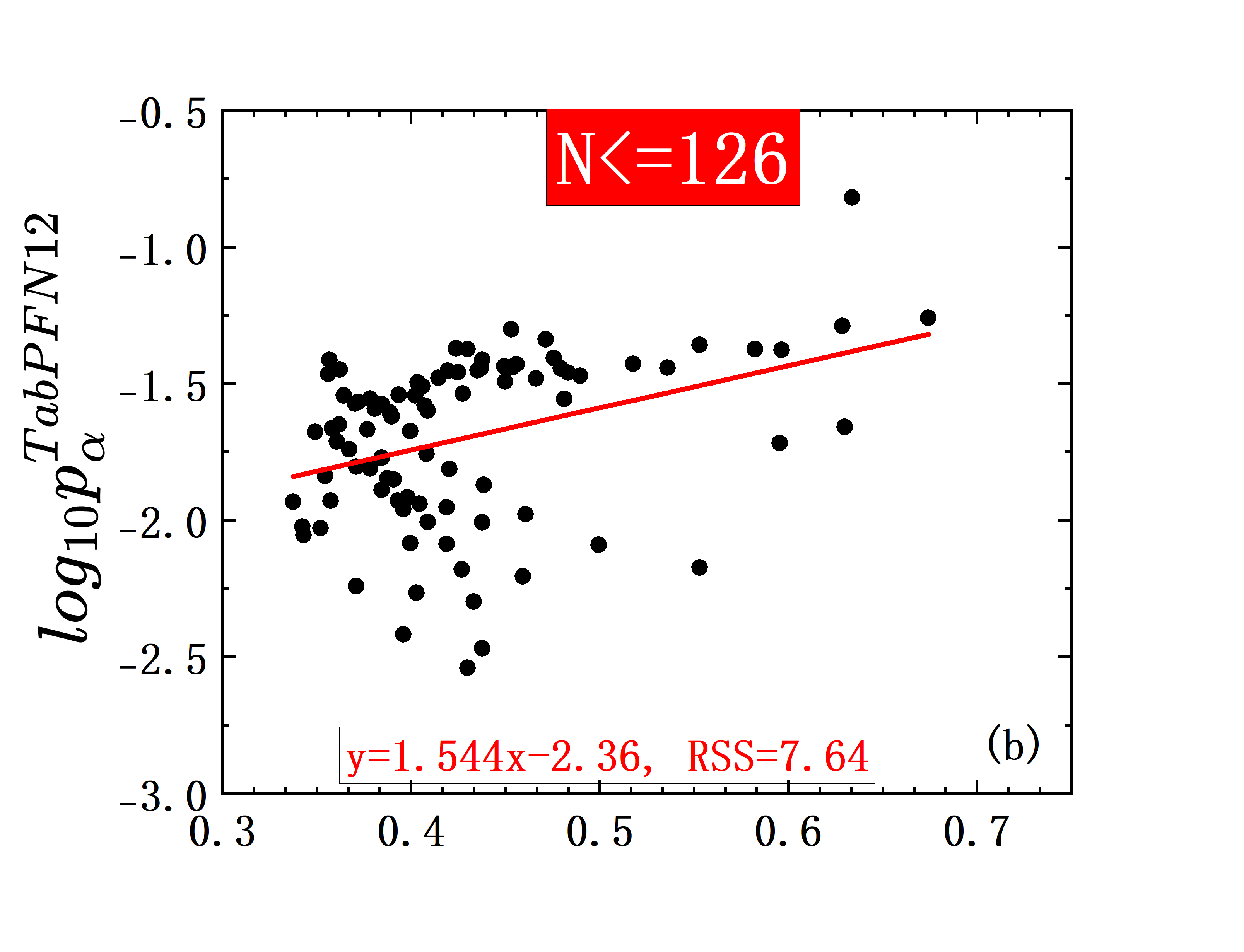}
  \end{subfigure}
  \hfill
  \begin{subfigure}{0.32\textwidth}
    \includegraphics[width=\linewidth]{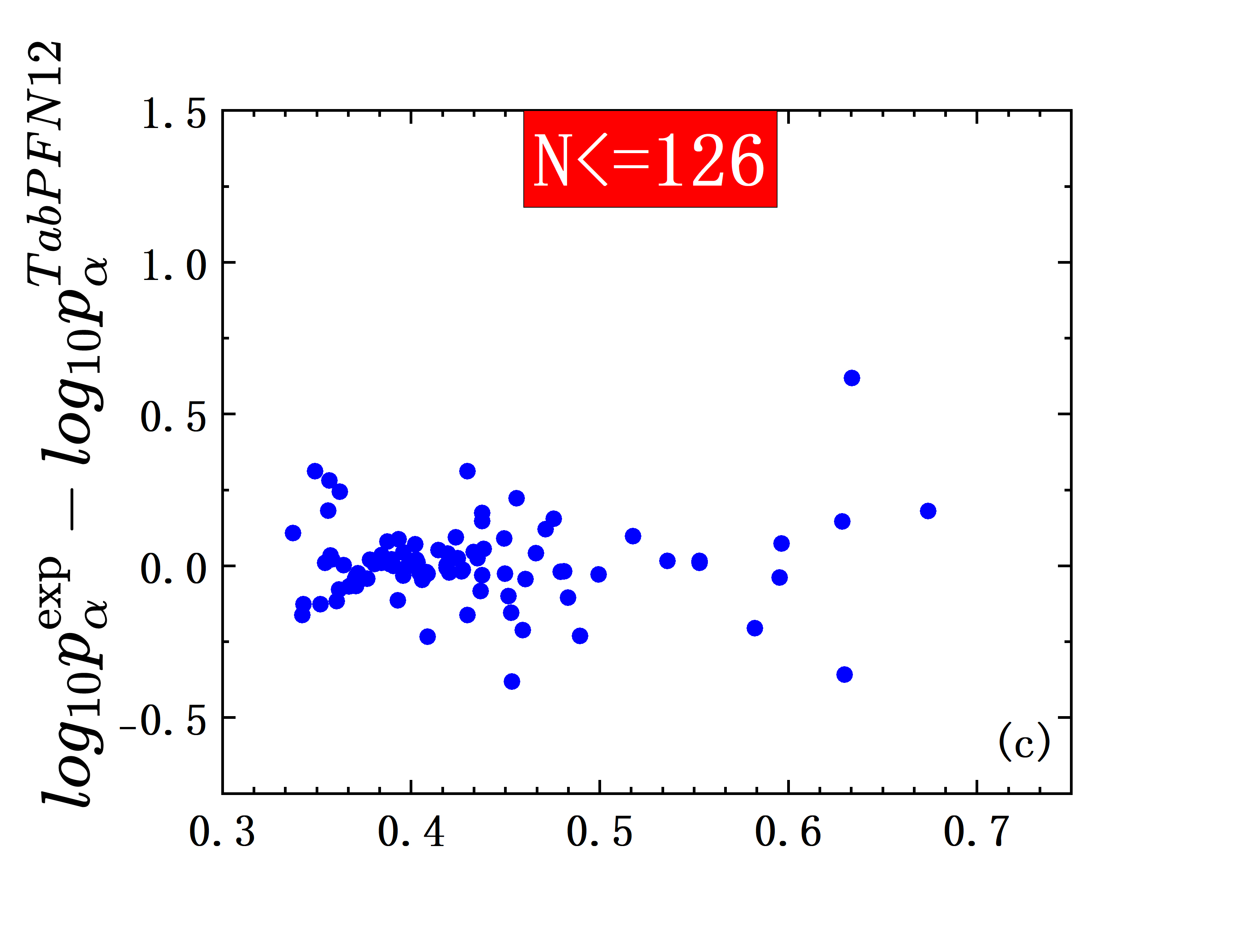}
  \end{subfigure}
  
  \vspace{-0.5cm}
  
  \begin{subfigure}{0.32\textwidth}
    \includegraphics[width=\linewidth]{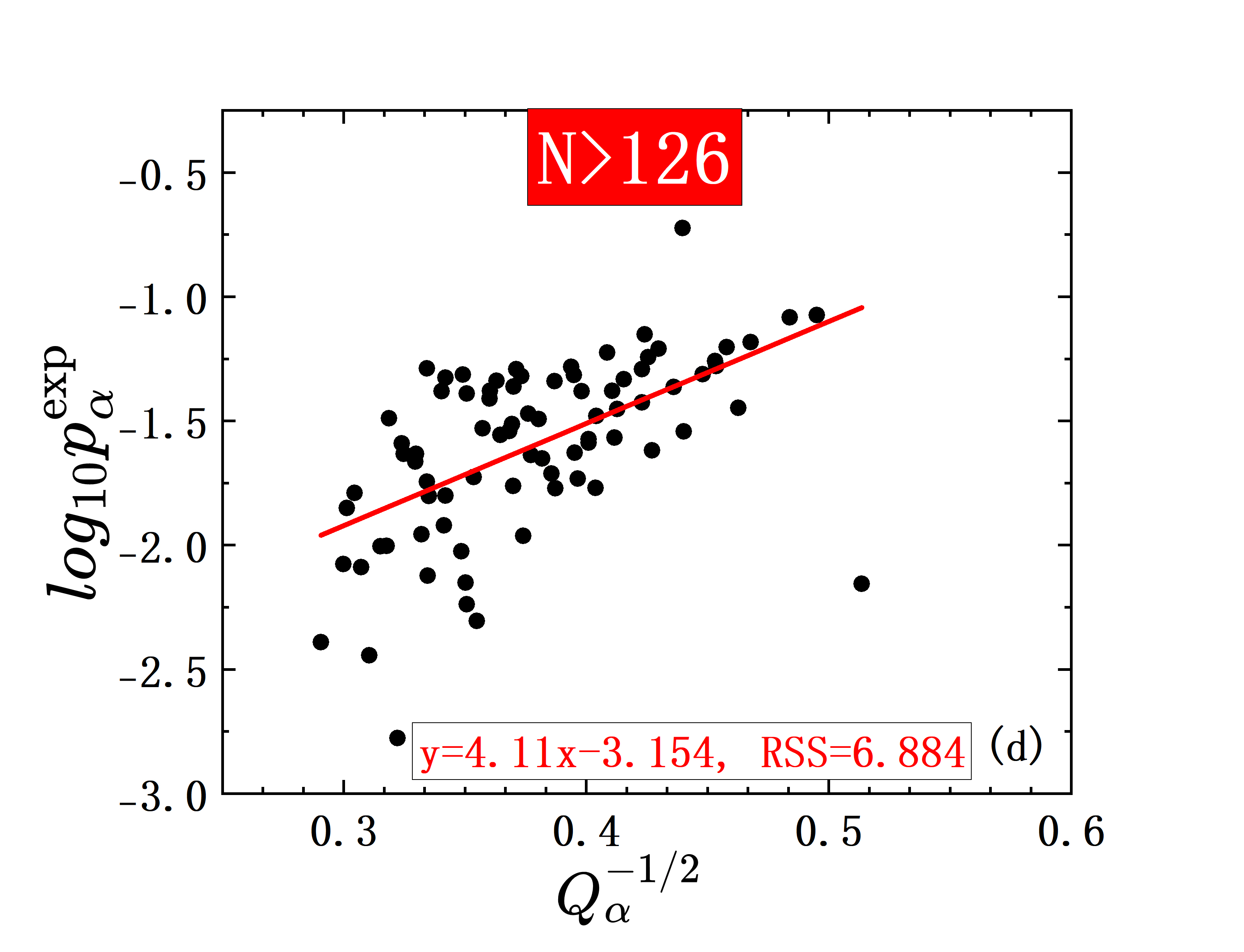}
  \end{subfigure}
  \hfill
  \begin{subfigure}{0.32\textwidth}
    \includegraphics[width=\linewidth]{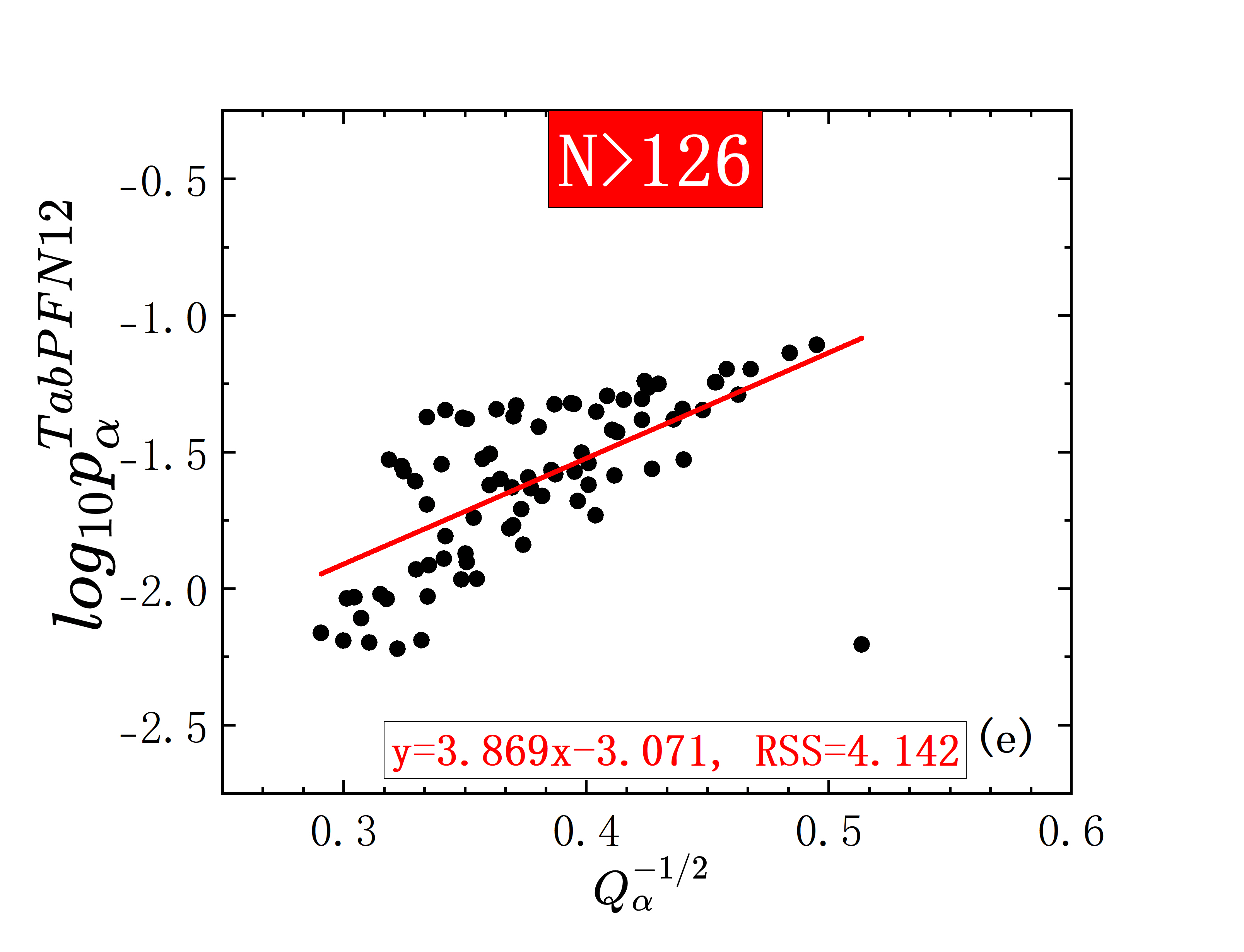}
  \end{subfigure}
  \hfill
  \begin{subfigure}{0.32\textwidth}
    \includegraphics[width=\linewidth]{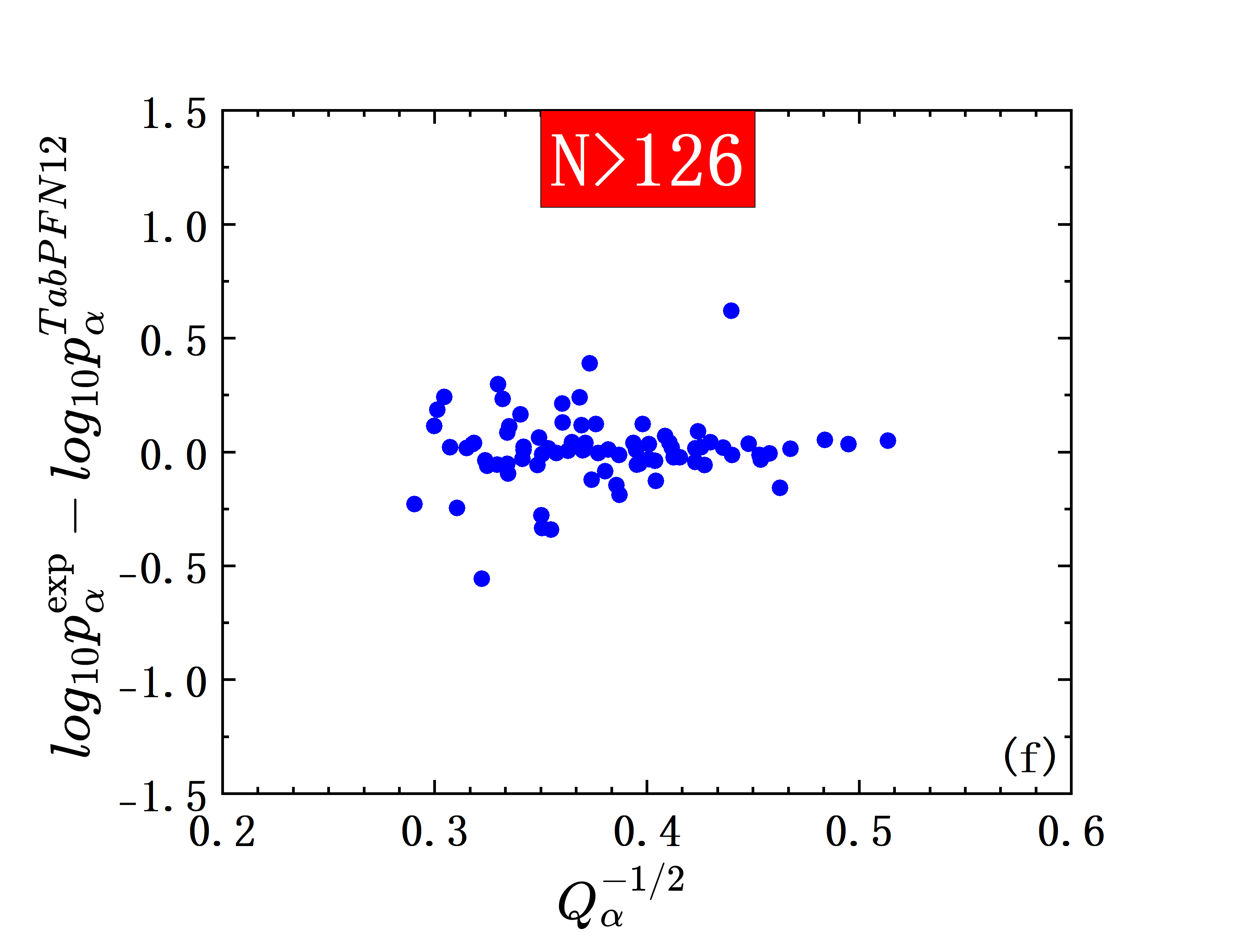}
  \end{subfigure}
  
  \caption{From left to right: the empirical logarithms of the $\alpha$-particle preformation factor $\log_{10} P_{\alpha}^{\mathrm{exp}}$, values predicted by the TabPFN12 method $\log_{10} P_{\alpha}^{\mathrm{TabPFN12}}$, and their logarithmic differences $\log_{10} P_{\alpha}^{\mathrm{exp}} - \log_{10} P_{\alpha}^{\mathrm{TabPFN12}}$, each plotted as functions of $Q_{\alpha}^{-1/2}$ (in MeV) for even–even nuclei. The upper row corresponds to nuclei with $N \leqslant 126$, and the lower row to $N > 126$. Linear regression results and residual sum of squares (RSS) values are provided to assess the quality of the fits.}
  \label{fig:alpha_preformation}
\end{figure*}

An approximately linear correlation is observed between the logarithm of the experimental $\alpha$-particle preformation factor, $\log_{10} P_{\alpha}^{\mathrm{exp}}$, and the inverse square root of the $\alpha$-decay energy, $Q_{\alpha}^{-1/2}$, in both the $N \leq 126$ and $N > 126$ regions \cite{deng2021correlation}. It is therefore pertinent to examine whether the TabPFN12 method employed in this study can adequately capture this empirical trend. Figures Figure~\ref{fig:alpha_preformation} (a) and ~\ref{fig:alpha_preformation} (b) in the upper row illustrate the relationship of $\log_{10} P_{\alpha}^{\mathrm{exp}}$ and $\log_{10} P_{\alpha}^{\mathrm{TabPFN12}}$ with $Q_{\alpha}^{-1/2}$ for nuclei with $N \leq 126$, while Figures~\ref{fig:alpha_preformation} (d) and~\ref{fig:alpha_preformation} (e) in the lower row show the corresponding results for the $N > 126$ region. Each plot includes a linear regression curve along with the residual sum of squares (RSS) to quantitatively assess the goodness of fit. Furthermore, Figs.~\ref{fig:alpha_preformation} (c) and~\ref{fig:alpha_preformation} (f) display the deviation between the experimental values and the TabPFN12 predictions, i.e., $\log_{10} P_{\alpha}^{\mathrm{exp}} - \log_{10} P_{\alpha}^{\mathrm{TabPFN12}}$, as a function of $Q_{\alpha}^{-1/2}$. The results confirm a clear linear dependence between $\log_{10} P_{\alpha}^{\mathrm{exp}}$ and $Q_{\alpha}^{-1/2}$, which is corroborated by the regression parameters. Moreover, the residuals are distributed randomly around zero without any systematic trend, indicating that the TabPFN model has effectively learned the underlying physical relationship between these quantities. These findings suggest that the Geiger-Nuttall law, traditionally applied to $\alpha$-decay half-lives, may also be extended to describe the $\alpha$-particle preformation factors.

\begin{figure}[htp]
  \centering
  
  \begin{subfigure}{0.5\textwidth}
    \centering
    \includegraphics[width=\linewidth, height=0.3\textheight, keepaspectratio]{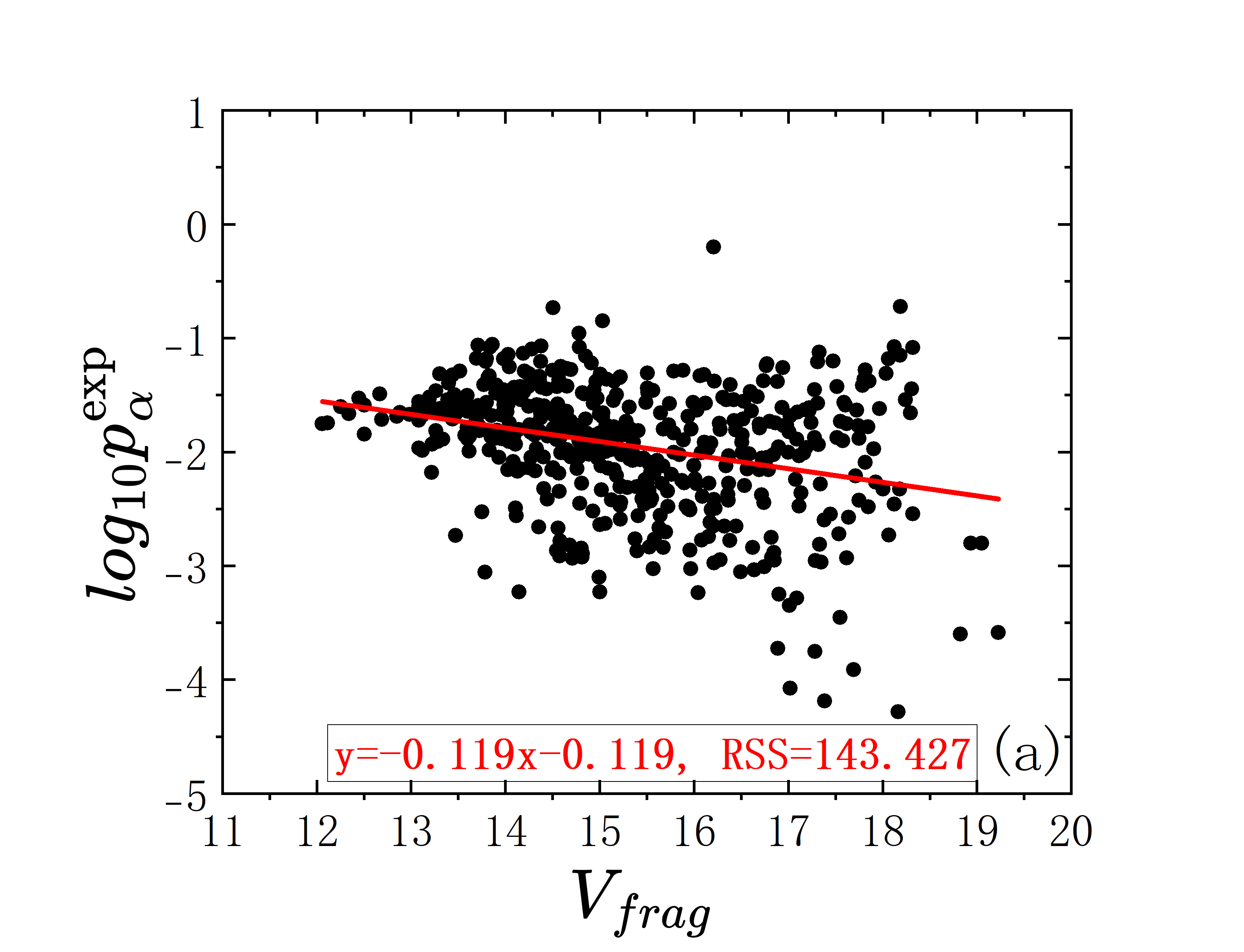}
  \end{subfigure}
  
  \begin{subfigure}{0.5\textwidth}
    \centering
    \includegraphics[width=\linewidth, height=0.3\textheight, keepaspectratio]{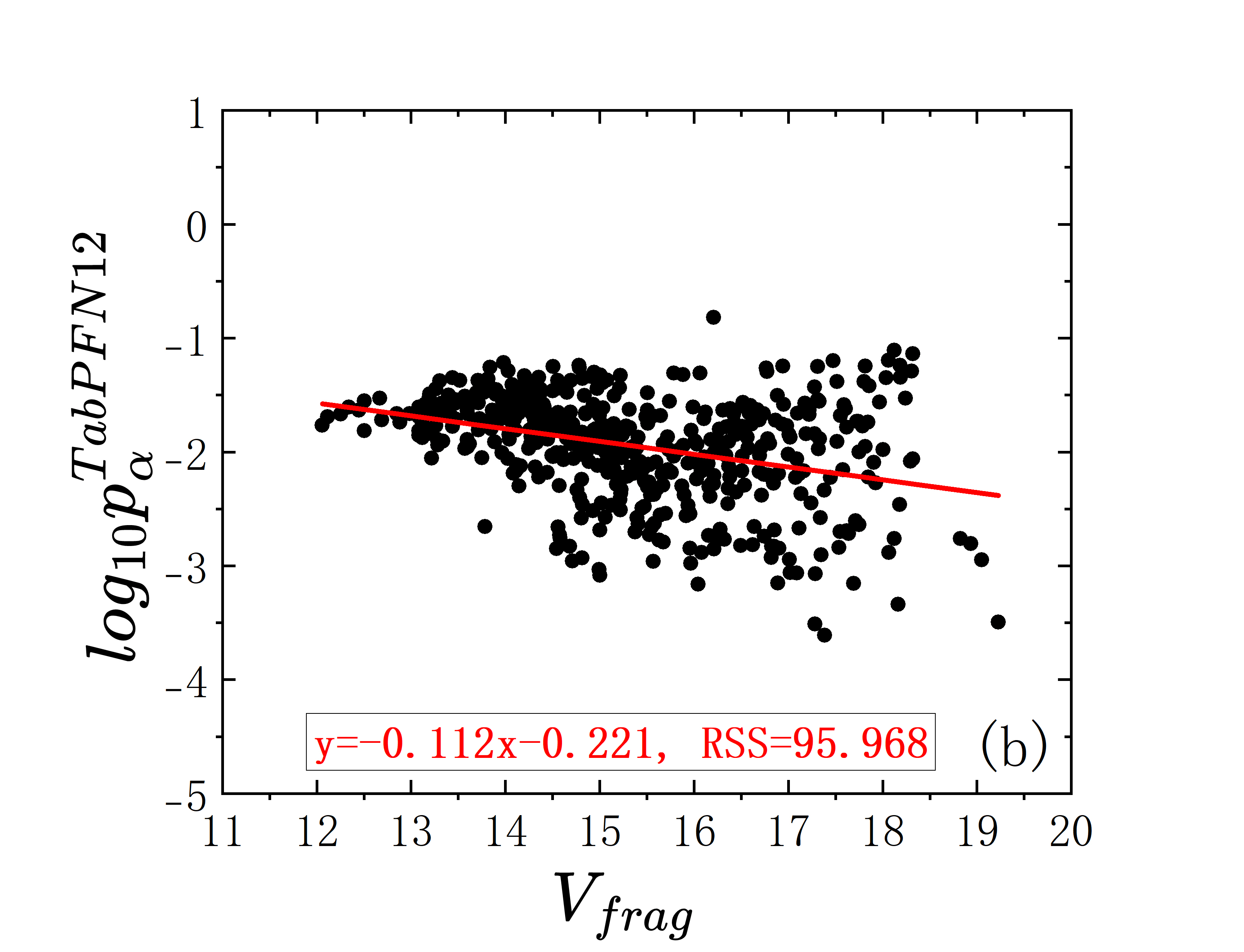}
  \end{subfigure}

  \caption{Correlation between the fragmentation potential $V_{\mathrm{frag}}$ and the logarithm of the $\alpha$-particle preformation factor for the 498 nuclei. (a) Experimentally extracted preformation factors $\log_{10}P_{\alpha}^{\mathrm{exp}}$. (b) Preformation factors predicted by the TabPFN12 model $\log_{10}P_{\alpha}^{\mathrm{TabPFN12}}$. The red lines represent linear fits to the data.}
  \label{fig:7}
\end{figure}

Besides the correlation with $Q_{\alpha}$, the $\alpha$-particle preformation probability is also expected to depend on the fragmentation potential $V_{\mathrm{frag}}$, which characterizes the energy cost of forming an $\alpha$ cluster in the parent nucleus \cite{poenaru2011heavy,delion2009universal}. To explore this dependence, the relationship between $\log_{10}P_{\alpha}$ and $V_{\mathrm{frag}}$ is examined for the same set of 498 nuclei. Figure~\ref{fig:7}(a) shows the correlation between the experimentally extracted preformation factors and the fragmentation potential. A clear negative linear trend is observed, indicating that larger fragmentation potentials correspond to smaller $\alpha$-cluster formation probabilities. The linear regression gives
\begin{equation}
\log_{10} P_{\alpha}^{\mathrm{exp}}
= -0.119 V_{\mathrm{frag}} - 0.119 .
\end{equation}

For comparison, Fig.~\ref{fig:7}(b) presents the corresponding relationship for the TabPFN12-predicted preformation factors. A very similar linear dependence is obtained,
\begin{equation}
\log_{10} P_{\alpha}^{\mathrm{TabPFN12}}
= -0.112 V_{\mathrm{frag}} - 0.221 .
\end{equation}

The close agreement between the slopes of the experimental and TabPFN12-derived correlations demonstrates that the machine learning model successfully captures the underlying physical dependence of the $\alpha$-cluster preformation probability on the fragmentation potential. Physically, a larger $V_{\mathrm{frag}}$ implies a higher energy cost for cluster formation, leading to a reduced probability for $\alpha$ preformation. This result further confirms that the TabPFN model not only reproduces experimental trends but also retains meaningful nuclear-structure information.

\begin{figure}
    \centering
    \includegraphics[width=1\linewidth]{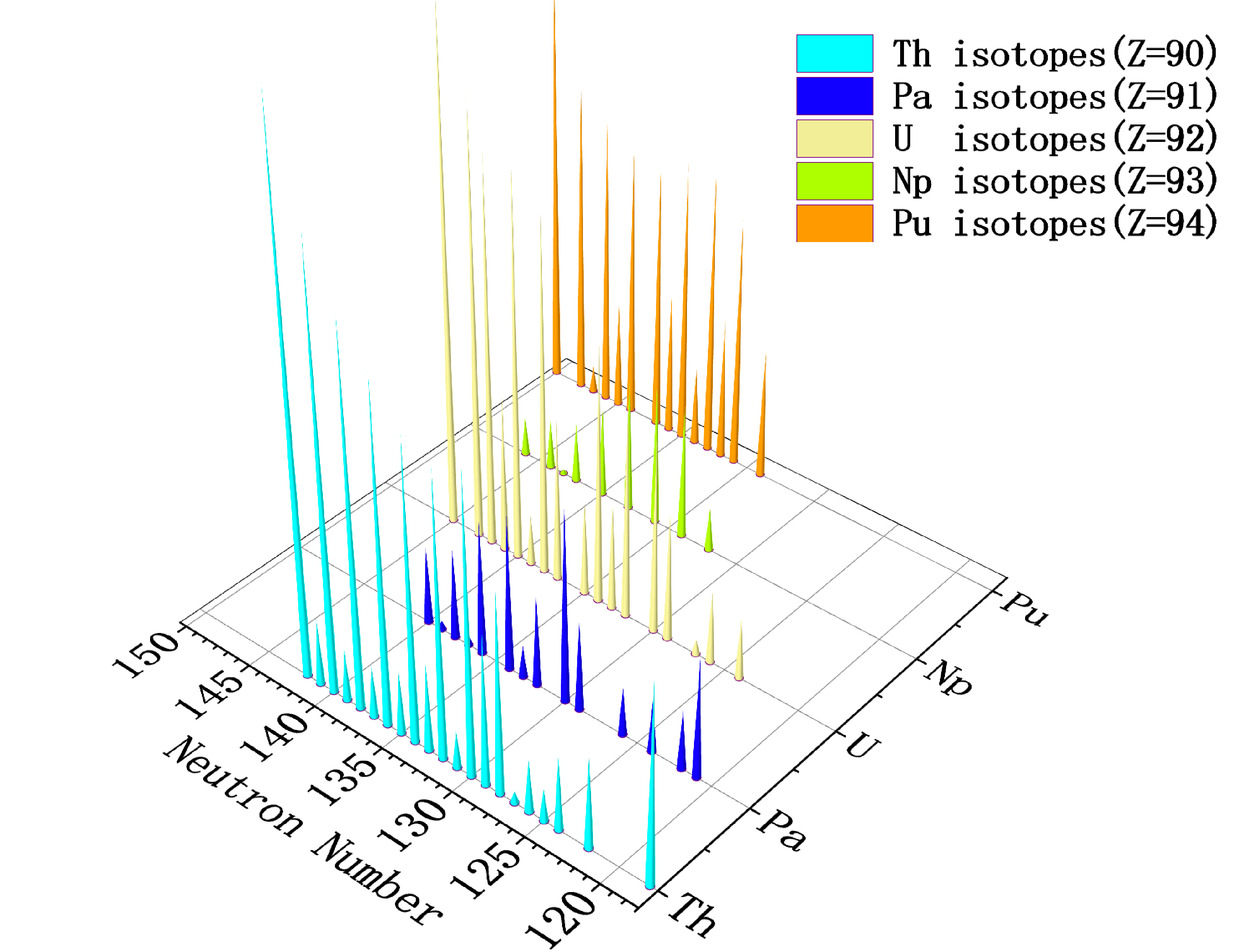}
    \caption{The $\alpha$-particle preformation factors computed using the TabPFN12 method, plotted as a function of neutron number, for the isotopic chains of Th, Pa, U, Np, and Pu.}
    \label{fig:3}
\end{figure}

The odd-even staggering (OES) effect is a well-established phenomenon manifested in various nuclear properties, including masses, nucleon separation energies, decay energies \cite{yang2022new}, and half-lives \cite{yang2022examining}. It is therefore of interest to investigate whether this effect also manifests in $\alpha$-particle preformation factors. Figure~\ref{fig:3} displays the preformation factors $P_{\alpha}^{\mathrm{TabPFN12}}$ computed using the TabPFN12 method for isotopic chains of Th, Pa, U, Np, and Pu. The results successfully reproduce the OES pattern. Specifically, for the even-$Z$ isotopic chains of Th, U, and Pu, $P_{\alpha}^{\mathrm{TabPFN12}}$ for odd-$N$ nuclei is systematically lower than for adjacent even-$N$ nuclei. This indicates that an unpaired neutron suppresses $\alpha$-cluster formation on the nuclear surface, thereby reducing $P_{\alpha}^{\mathrm{TabPFN12}}$. Similarly, comparing the odd-$Z$ Pa chain with the even-$Z$ Th chain reveals that, for the same neutron number $N$, $P_{\alpha}^{\mathrm{TabPFN12}}$ values for Pa isotopes are generally smaller, implying that an unpaired proton also hinders $\alpha$-particle formation. More systematically, the Th chain ($Z = 90$) exhibits larger $P_{\alpha}^{\mathrm{TabPFN12}}$ values than the adjacent Pa chain ($Z = 91$), and the Pu chain ($Z = 94$) shows higher values than the neighboring Np chain ($Z = 93$). At $N = 144$, the preformation factor of U isotopes exceeds those of both the Th and Pu chains. These systematic trends, evident in Fig.~\ref{fig:3}, indicate that the $\alpha$-preformation factor increases with proton number after crossing the magic number $Z = 82$, reaches a maximum around $Z = 92$, and gradually decreases thereafter. The presence of unpaired nucleons reduces the probability of $\alpha$-cluster formation on the nuclear surface. Consequently, the behavior and evolution of $\alpha$-preformation factors provide valuable insights into nuclear structure.

Previous studies have commonly derived $\alpha$-particle preformation factors from empirical formulas. For comparison with the present TabPFN12-based results, the empirical formulations from Refs. \cite{deng2021correlation} and \cite{he2021improved} are adopted. The corresponding root-mean-square deviations $\sigma_{\mathrm{rms}}$ are summarized in Table II. A comparison demonstrates that the machine learning approach delivers performance comparable to or superior to that of empirical formulas. Specifically, relative to Ref. \cite{he2021improved}, the $\sigma_{\mathrm{rms}}$ values for the four nuclear categories decrease from 0.544, 0.632, 0.564, and 0.610 to 0.153, 0.254, 0.220, and 0.249, respectively. When compared to Ref. \cite{deng2021correlation}, the corresponding $\sigma_{\mathrm{rms}}$ values are reduced from 0.548, 0.713, 0.655, and 0.646 to the same set of improved values. The overall $\sigma_{\mathrm{rms}}$ is reduced from 0.584 and 0.637 to 0.211, corresponding to improvements of 63.8\% and 66.8\%, respectively.

\begin{table}
\caption{A comparison of the $\sigma_{\text{rms}}$ values between the improved TabPFN12 model and empirical approaches.}
\begin{tabular}{lccc}
\hline
nuclei type      & TabPFN12 & \multicolumn{1}{l}{ref. \cite{he2021improved}} & ref. \cite{deng2021correlation} \\ \hline
even-even nuclei & 0.153   & 0.544                                                          & 0.548                                            \\
even-odd nuclei  & 0.254  & 0.632                                                          & 0.713                                            \\
odd-even nuclei  & 0.22  & 0.564                                                          & 0.655                                            \\
odd-odd nuclei   & 0.249  & 0.61                                                           & 0.646                                            \\
all              & 0.211  & 0.584                                                          & 0.637                                            \\ \hline
\end{tabular}
\end{table}

\subsection{Extrapolation capability of TabPFN12 neural network}

The $\alpha$-decay half-life is an experimentally measurable quantity. Theoretical half-lives incorporating the $\alpha$-particle preformation probability are expected to show improved agreement with experimental data. To assess the performance of the TabPFN12-derived preformation factors, we compute them for 41 newly evaluated nuclei with $Z>104$ in NUBASE2020—an update from the 2016 evaluation—and compare them with predictions from the empirical formula of Ref. \cite{deng2021correlation}. The corresponding $\alpha$-decay energies are taken from the AME2020 atomic mass evaluation. Detailed results are summarized in Table III. The $P_\alpha$ values from the TabPFN12 approach are largely consistent with the empirical systematics, indicating that the model can reliably extrapolate preformation factors to unexplored regions of the nuclear chart.

\begin{figure}[htbp]
  \centering
  
  \begin{subfigure}{0.5\textwidth}
    \centering
    \includegraphics[width=\linewidth, height=0.3\textheight, keepaspectratio]{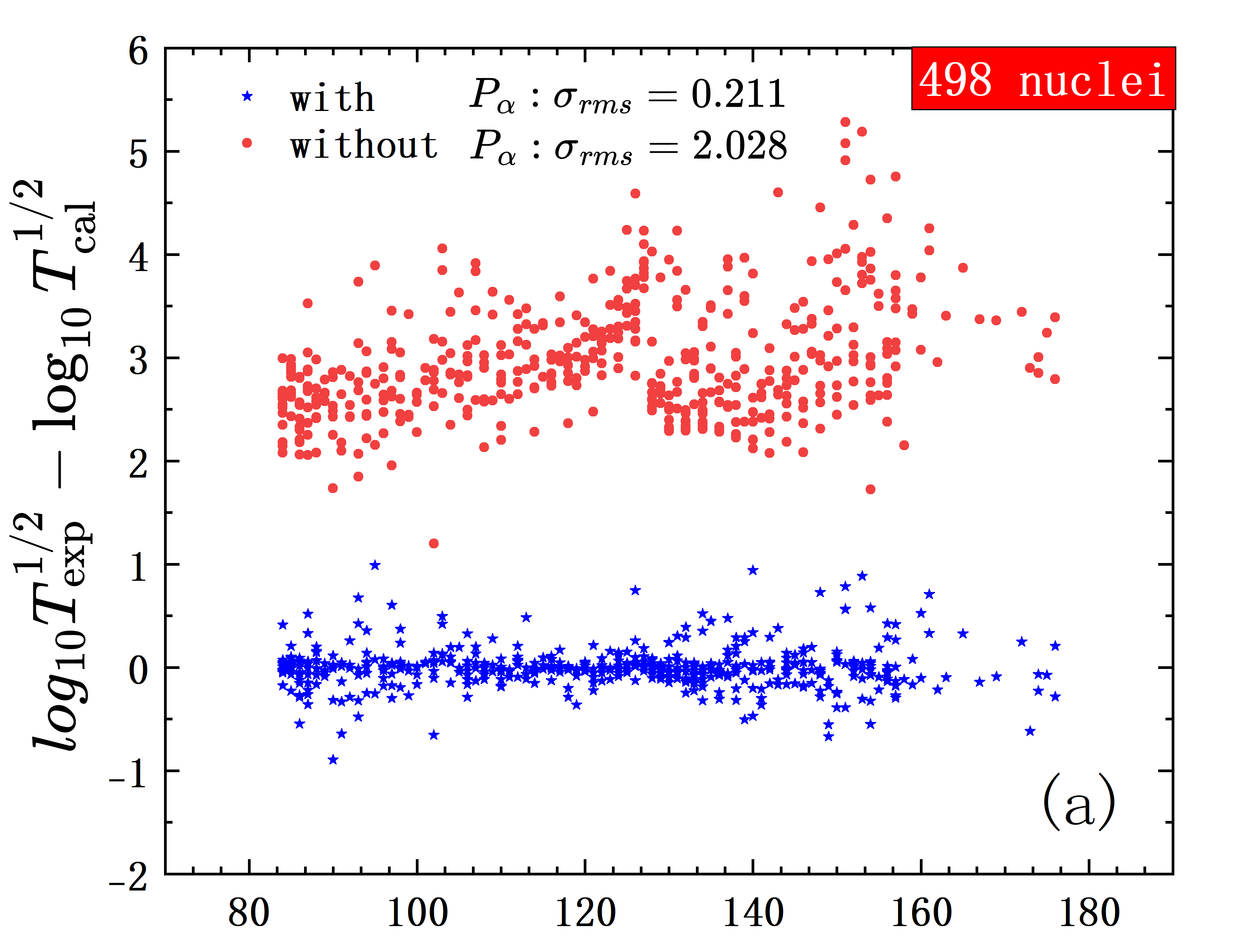}
  \end{subfigure}
  
  \vspace{10pt}  
  
  \begin{subfigure}{0.5\textwidth}
    \centering
    \includegraphics[width=\linewidth, height=0.3\textheight, keepaspectratio]{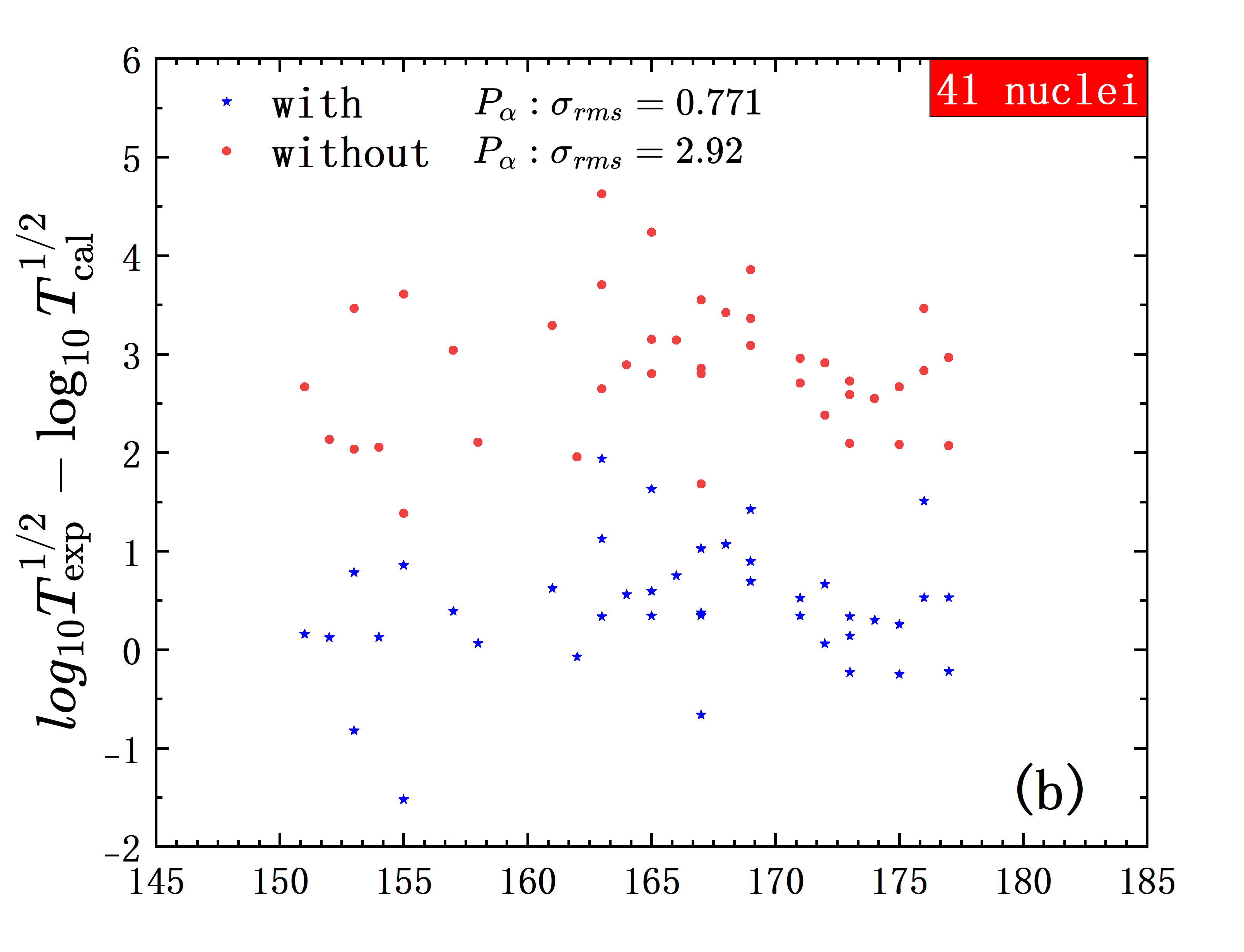}
  \end{subfigure}

  \begin{subfigure}{0.5\textwidth}
    \centering
    \includegraphics[width=\linewidth, height=0.3\textheight, keepaspectratio]{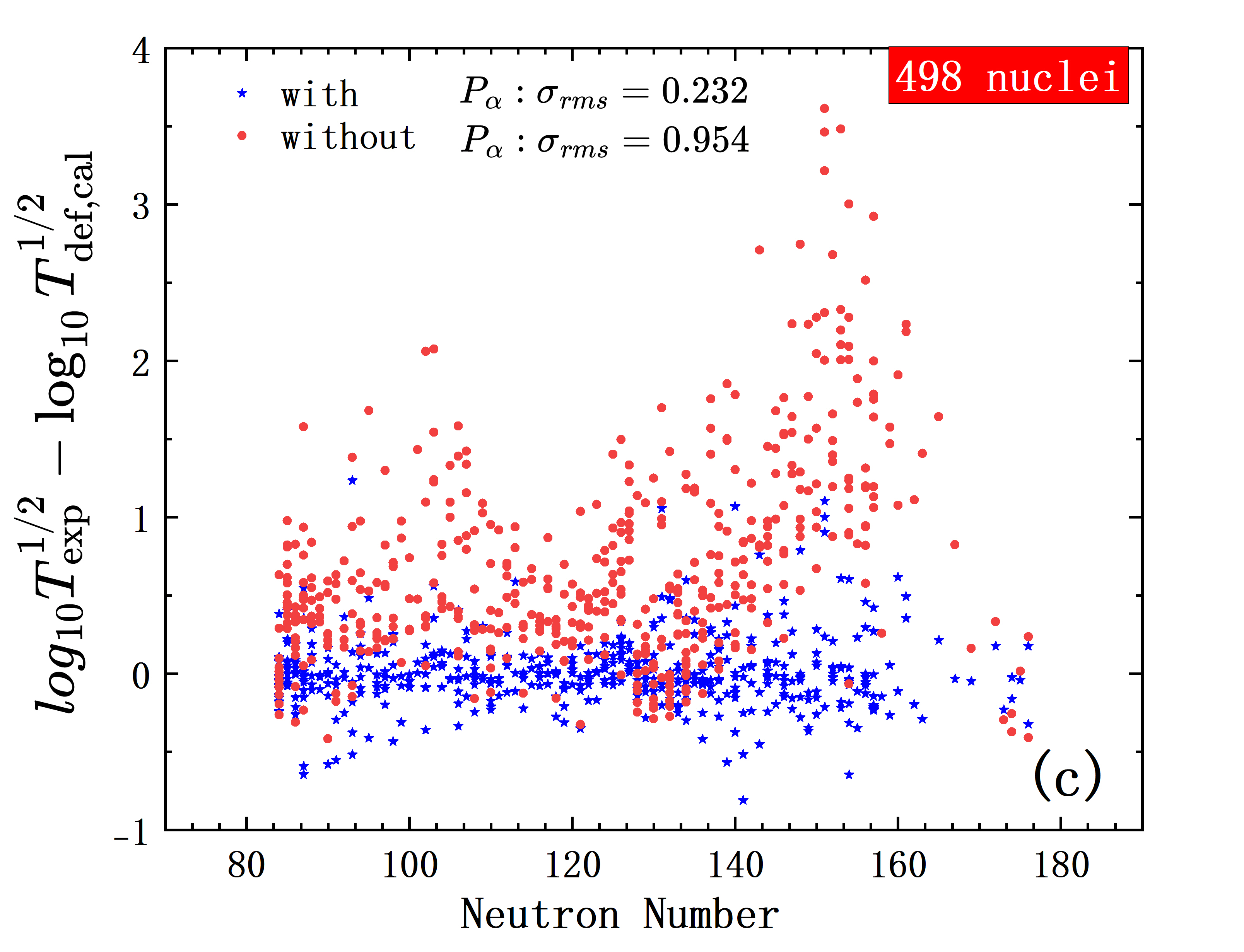}
  \end{subfigure}

  \caption{Deviations between experimental and calculated $\alpha$-decay half-lives. (a) Results for 498 nuclei obtained within the CPPM framework. (b) Predictions for an additional set of 41 nuclei. (c) Results for 498 nuclei calculated using the deformed Coulomb plus Woods–Saxon potential. Blue stars denote calculations including the TabPFN12 preformation factor, while red circles correspond to results without the preformation factor.}
  \label{fig:4}
\end{figure}

Furthermore, $\alpha$-decay half-lives are calculated by incorporating the TabPFN12 preformation factors and compared with experimental data. Figure~\ref{fig:4} shows the deviations $\log_{10}(T^{\mathrm{exp}}_{1/2})-\log_{10}(T^{\mathrm{cal}}_{1/2})$ between experimental and calculated half-lives. Figure~\ref{fig:4} (a) presents the results for 498 nuclei within the CPPM framework. Without including the preformation factor, the calculated half-lives show large systematic deviations from the experimental values, yielding a root-mean-square deviation $\sigma_{\mathrm{rms}}=2.028$. After incorporating the TabPFN12 preformation factor, the deviations become concentrated around zero and the rms deviation is significantly reduced to $\sigma_{\mathrm{rms}}=0.211$, corresponding to an improvement of about 89.5\%. To further test the predictive capability of the model, the same procedure is applied to an additional set of 41 nuclei, as shown in Fig.~\ref{fig:4} (b). A similar trend is observed: the rms deviation decreases from 2.920 to 0.771 when the TabPFN12 preformation factor is included, indicating that the model retains good predictive performance for nuclei outside the training dataset. For comparison, Fig.~\ref{fig:4} (c) shows the deviations obtained using a deformed Coulomb plus Woods–Saxon potential for the same 498 nuclei. The inclusion of the TabPFN12 preformation factor again significantly reduces the deviations, with the rms value decreasing from 0.954 to 0.232. Although this value is slightly larger than that obtained within the CPPM framework ($\sigma_{\mathrm{rms}}=0.211$), the calculated half-lives remain well clustered around the experimental values. This indicates that the TabPFN12-derived preformation factors provide a robust improvement to $\alpha$-decay half-life predictions for both spherical and deformed potential descriptions.

\begin{table*} 
\caption{A comparison of $\alpha$-particle preformation factors between TabPFN predictions and empirical formulations from Refs. \cite{deng2021correlation,he2021improved} is presented for 41 newly evaluated nuclei in the NUBASE2020 database \cite{kondev2021nubase2020}. The quantities $P^{\text{Deng}}_{\alpha}$ and $P^{\text{He}}_{\alpha}$ correspond to preformation factors calculated using the empirical relations in Refs. \cite{deng2021correlation} and \cite{he2021improved}, respectively, while $P^{\text{TabPFN12}}_{\alpha}$ denotes the values predicted by the TabPFN12 model.}
\centering
\small 
\setlength{\tabcolsep}{3pt} 
\begin{tabular}{@{}c c c c c c c c c@{}} 
\hline
$\alpha$ transition & $Q_{\alpha}$ & $l_{\min}$ & $\log_{10}P^{\text{Deng}}_{\alpha}$ & $\log_{10}P^{\text{He}}_{\alpha}$ & $\log_{10}P^{\text{TabPFN12}}_{\alpha}$ & $\log_{10}T^{\text{exp}}_{1/2}$ & $\log_{10}T^{\text{CPPM}}_{1/2}$ & $\log_{10} T_{1/2}^{\mathrm{CPPM} + P_\alpha}$ \\ 
\hline
$^{255}\text{Rf}\rightarrow^{251}\text{No}$ & 9.06 & 1 & -1.93 & -2.13 & -2.51 & 0.49 & -2.18 & 0.33 \\ 
$^{256}\text{Rf}\rightarrow^{252}\text{No}$ & 8.93 & 0 & -1.74 & -2.62 & -2.01 & 0.33 & -1.09 & 0.21 \\ 
$^{257}\text{Rf}\rightarrow^{253}\text{No}$ & 9.08 & 5 & -2.74 & -2.50 & -2.86 & 0.75 & -1.28 & 1.57 \\ 
$^{258}\text{Rf}\rightarrow^{254}\text{No}$ & 9.20 & 0 & -1.76 & -1.71 & -1.93 & 0.59 & -2.64 & -0.72 \\ 
$^{269}\text{Sg}\rightarrow^{265}\text{Rf}$ & 8.58 & 0 & -2.27 & -2.03 & -2.31 & 2.48 & -0.17 & 2.14 \\ 
$^{260}\text{Bh}\rightarrow^{256}\text{Db}$ & 10.40 & 0 & -2.72 & -2.73 & -2.68 & -1.39 & -4.83 & -2.17 \\ 
$^{262}\text{Bh}\rightarrow^{258}\text{Db}$ & 10.32 & 0 & -2.77 & -2.74 & -2.75 & -1.38 & -4.69 & -1.94 \\ 
$^{270}\text{Bh}\rightarrow^{266}\text{Db}$ & 9.06 & 0 & -2.90 & -2.64 & -2.58 & 2.36 & -1.34 & 1.24 \\ 
$^{271}\text{Bh}\rightarrow^{267}\text{Db}$ & 9.42 & 0 & -2.29 & -2.16 & -2.34 & 0.46 & -2.43 & -0.10 \\ 
$^{272}\text{Bh}\rightarrow^{268}\text{Db}$ & 9.30 & 0 & -2.90 & -2.68 & -2.56 & 1.05 & -2.27 & 0.46 \\ 
$^{274}\text{Bh}\rightarrow^{270}\text{Db}$ & 8.94 & 0 & -2.90 & -2.69 & -2.43 & 1.76 & -1.04 & 1.39 \\ 
$^{263}\text{Hs}\rightarrow^{259}\text{Sg}$ & 10.73 & 5 & -2.90 & -2.74 & -2.91 & -3.05 & -4.43 & -1.53 \\ 
$^{266}\text{Hs}\rightarrow^{262}\text{Sg}$ & 10.35 & 0 & -1.97 & -1.92 & -2.04 & 2.34 & -4.50 & -2.47 \\ 
$^{270}\text{Hs}\rightarrow^{266}\text{Sg}$ & 9.07 & 0 & -2.50 & -1.80 & -2.03 & 0.95 & -1.01 & 1.02 \\ 
$^{273}\text{Hs}\rightarrow^{269}\text{Sg}$ & 9.65 & 0 & -2.32 & -2.20 & -2.46 & 0.03 & -2.77 & -0.31 \\ 
$^{275}\text{Hs}\rightarrow^{271}\text{Sg}$ & 9.45 & 0 & -2.32 & -2.19 & -2.34 & -0.55 & -2.23 & 0.11 \\ 
$^{266}\text{Mt}\rightarrow^{262}\text{Bh}$ & 11.00 & 0 & -2.83 & -2.82 & -2.65 & -2.70 & -5.74 & -3.09 \\ 
$^{275}\text{Mt}\rightarrow^{271}\text{Bh}$ & 10.48 & 0 & -2.35 & -2.31 & -2.39 & -1.51 & -4.61 & -2.26 \\ 
$^{276}\text{Mt}\rightarrow^{272}\text{Bh}$ & 10.10 & 0 & -2.95 & -2.81 & -2.52 & 0.15 & -3.70 & -1.17 \\ 
$^{278}\text{Mt}\rightarrow^{274}\text{Bh}$ & 9.58 & 0 & -2.98 & -2.76 & -2.39 & 0.78 & -2.30 & 0.09 \\ 
$^{272}\text{Rg}\rightarrow^{268}\text{Mt}$ & 11.20 & 0 & -2.95 & -2.92 & -2.67 & -2.38 & -5.67 & -3.00 \\ 
$^{274}\text{Rg}\rightarrow^{270}\text{Mt}$ & 11.48 & 0 & -2.98 & -2.95 & -2.69 & -1.60 & -6.63 & -3.64 \\ 
$^{278}\text{Rg}\rightarrow^{274}\text{Mt}$ & 10.85 & 0 & -2.99 & -2.94 & -2.51 & -2.10 & -4.96 & -2.45 \\ 
$^{279}\text{Rg}\rightarrow^{275}\text{Mt}$ & 10.53 & 0 & -2.42 & -2.36 & -2.35 & 0.77 & -4.19 & -1.84 \\ 
$^{280}\text{Rg}\rightarrow^{276}\text{Mt}$ & 10.15 & 0 & -3.02 & -2.85 & -2.43 & 0.04 & -3.22 & -0.79 \\ 
$^{278}\text{Nh}\rightarrow^{274}\text{Rg}$ & 11.99 & 0 & -3.00 & -3.03 & -2.61 & -2.64 & -6.88 & -4.27 \\ 
$^{282}\text{Nh}\rightarrow^{278}\text{Rg}$ & 10.78 & 0 & -3.05 & -2.95 & -2.47 & 0.85 & -4.21 & -1.75 \\ 
$^{284}\text{Nh}\rightarrow^{280}\text{Rg}$ & 10.28 & 0 & -3.07 & -2.91 & -2.43 & 0.01 & -2.97 & -0.53 \\ 
$^{285}\text{Nh}\rightarrow^{281}\text{Rg}$ & 10.01 & 0 & -2.52 & -2.36 & -2.24 & 0.66 & -2.25 & -0.01 \\ 
$^{286}\text{Nh}\rightarrow^{282}\text{Rg}$ & 9.79 & 0 & -3.13 & -2.57 & -2.39 & 1.08 & -1.64 & 0.74 \\ 
$^{285}\text{Fl}\rightarrow^{281}\text{Cn}$ & 10.56 & 0 & -2.52 & -2.43 & -2.36 & 0.08 & -3.38 & -1.02 \\ 
$^{287}\text{Fl}\rightarrow^{283}\text{Cn}$ & 10.17 & 0 & -2.55 & -2.39 & -2.32 & 0.29 & -2.38 & -0.06 \\ 
$^{290}\text{Fl}\rightarrow^{286}\text{Cn}$ & 9.86 & 0 & -2.20 & -2.05 & -1.96 & 1.90 & -1.62 & 0.39 \\ 
$^{287}\text{Mc}\rightarrow^{283}\text{Nh}$ & 10.76 & 0 & -2.56 & -2.47 & -2.32 & -1.22 & -3.60 & -1.28 \\ 
$^{288}\text{Mc}\rightarrow^{284}\text{Nh}$ & 10.65 & 0 & -3.15 & -2.99 & -2.45 & -1.70 & -3.34 & -0.89 \\ 
$^{289}\text{Mc}\rightarrow^{285}\text{Nh}$ & 10.49 & 0 & -2.58 & -2.45 & -2.25 & -0.39 & -2.96 & -0.69 \\ 
$^{290}\text{Mc}\rightarrow^{286}\text{Nh}$ & 10.41 & 0 & -3.19 & -2.92 & -2.41 & 0.08 & -2.75 & -0.34 \\ 
$^{291}\text{Lv}\rightarrow^{287}\text{Fl}$ & 10.89 & 0 & -2.61 & -2.51 & -2.33 & -1.59 & -3.67 & -1.34 \\ 
$^{293}\text{Lv}\rightarrow^{289}\text{Fl}$ & 10.68 & 0 & -2.63 & -2.50 & -2.29 & -1.10 & -3.17 & -0.88 \\ 
$^{293}\text{Ts}\rightarrow^{289}\text{Mc}$ & 11.32 & 0 & -2.64 & -2.57 & -2.31 & -1.60 & -4.43 & -2.13 \\ 
$^{294}\text{Ts}\rightarrow^{290}\text{Mc}$ & 11.18 & 0 & -3.24 & -3.09 & -2.43 & -1.15 & -4.11 & -1.68 \\ 
\hline
\end{tabular}  
\end{table*}

\begin{table*}[!t]
\centering
\caption{Predicted $\alpha$-decay half-lives for superheavy nuclei with $Z=117-120$, where decay energies (in MeV) are adopted from Ref. \cite{ma2019basic}. Half-lives are presented in units of seconds.}
\resizebox{\linewidth}{!}{
\begin{tabular}{cccccccccccccc}
\hline
\textbf{$\alpha$ transition} & \textbf{\begin{tabular}[c]{@{}c@{}}$Q_{\alpha}$\end{tabular}} & $l_{\min}$ & $\log_{10} T_{1/2}^{\mathrm{CPPM} + P_\alpha}$ & \textbf{UDL} & \textbf{Ni} & \textbf{Royer} & \textbf{$\alpha$ transition} & \textbf{\begin{tabular}[c]{@{}c@{}}$Q_{\alpha}$\end{tabular}} & $l_{\min}$ & $\log_{10} T_{1/2}^{\mathrm{CPPM} + P_\alpha}$ & \textbf{UDL} & \textbf{Ni} & \textbf{Royer} \\ \hline
\textsuperscript{279}Ts → \textsuperscript{275}Mc & 13.49 & 2 & -6.04 & -7.17 & -6.17 & -6.55 & \textsuperscript{315}Og → \textsuperscript{311}Lv & 8.40 & 3 & 6.46 & 5.86 & 6.64 & 6.04 \\
\textsuperscript{281}Ts → \textsuperscript{277}Mc & 13.52 & 2 & -6.13 & -7.26 & -6.22 & -6.64 & \textsuperscript{316}Og → \textsuperscript{312}Lv & 8.47 & 0 & 8.79 & 7.34 & 8.86 & 8.30 \\
\textsuperscript{283}Ts → \textsuperscript{279}Mc & 13.00 & 2 & -5.24 & -6.30 & -5.24 & -5.68 & \textsuperscript{317}Og → \textsuperscript{313}Lv & 8.09 & 3 & 6.37 & 5.79 & 6.61 & 5.97 \\
\textsuperscript{285}Ts → \textsuperscript{281}Mc & 12.39 & 2 & -4.10 & -5.08 & -4.01 & -4.46 & \textsuperscript{318}Og → \textsuperscript{314}Lv & 8.48 & 0 & -5.90 & -7.02 & -5.94 & -6.37 \\
\textsuperscript{287}Ts → \textsuperscript{283}Mc & 11.92 & 2 & -3.16 & -4.08 & -2.99 & -3.46 & \textsuperscript{285}119 → \textsuperscript{281}Ts & 13.65 & 2 & -5.33 & -6.35 & -5.24 & -5.70 \\
\textsuperscript{289}Ts → \textsuperscript{285}Mc & 11.79 & 2 & -2.93 & -3.82 & -2.70 & -3.20 & \textsuperscript{287}119 → \textsuperscript{283}Ts & 13.28 & 2 & -4.93 & -5.99 & -4.86 & -5.34 \\
\textsuperscript{291}Ts → \textsuperscript{287}Mc & 11.46 & 2 & -2.23 & -3.08 & -1.94 & -2.46 & \textsuperscript{289}119 → \textsuperscript{285}Ts & 13.08 & 2 & -4.58 & -5.60 & -4.44 & -4.95 \\
\textsuperscript{293}Ts → \textsuperscript{289}Mc & 11.37 & 2 & -2.08 & -2.90 & -1.73 & -2.28 & \textsuperscript{291}119 → \textsuperscript{287}Ts & 12.87 & 2 & -3.90 & -4.88 & -3.70 & -4.23 \\
\textsuperscript{295}Ts → \textsuperscript{291}Mc & 11.12 & 2 & -1.51 & -2.32 & -1.12 & -1.70 & \textsuperscript{293}119 → \textsuperscript{289}Ts & 12.51 & 2 & -4.04 & -5.04 & -3.82 & -4.40 \\
\textsuperscript{297}Ts → \textsuperscript{293}Mc & 11.54 & 2 & -2.49 & -3.37 & -2.13 & -2.76 & \textsuperscript{295}119 → \textsuperscript{291}Ts & 12.57 & 2 & -3.55 & -4.49 & -3.25 & -3.85 \\
\textsuperscript{299}Ts → \textsuperscript{295}Mc & 11.39 & 2 & -2.17 & -3.05 & -1.77 & -2.44 & \textsuperscript{297}119 → \textsuperscript{293}Ts & 12.30 & 2 & -4.41 & -5.44 & -4.15 & -4.81 \\
\textsuperscript{301}Ts → \textsuperscript{297}Mc & 11.54 & 2 & -2.54 & -3.44 & -2.13 & -2.84 & \textsuperscript{299}119 → \textsuperscript{295}Ts & 12.73 & 2 & -3.94 & -4.71 & -3.40 & -4.07 \\
\textsuperscript{303}Ts → \textsuperscript{299}Mc & 12.73 & 2 & -5.04 & -6.09 & -4.70 & -5.50 & \textsuperscript{301}119 → \textsuperscript{297}Ts & 12.37 & 0 & -3.75 & -4.74 & -3.40 & -4.11 \\
\textsuperscript{305}Ts → \textsuperscript{301}Mc & 12.13 & 2 & -3.86 & -4.85 & -3.45 & -4.26 & \textsuperscript{303}119 → \textsuperscript{299}Ts & 12.37 & 2 & -5.84 & -6.98 & -5.56 & -6.35 \\
\textsuperscript{307}Ts → \textsuperscript{303}Mc & 11.00 & 1 & -1.44 & -2.21 & -0.82 & -1.61 & \textsuperscript{305}119 → \textsuperscript{301}Ts & 13.45 & 2 & -4.94 & -5.76 & -4.33 & -5.13 \\
\textsuperscript{309}Ts → \textsuperscript{305}Mc & 10.01 & 3 & 1.44 & 0.46 & 1.84 & 1.07 & \textsuperscript{307}119 → \textsuperscript{303}Ts & 12.82 & 0 & -1.34 & -2.46 & -1.05 & -1.82 \\
\textsuperscript{311}Ts → \textsuperscript{307}Mc & 8.84 & 2 & 4.98 & 4.19 & 5.54 & 4.81 & \textsuperscript{309}119 → \textsuperscript{305}Ts & 11.34 & 3 & 0.04 & -0.56 & 0.84 & 0.07 \\
\textsuperscript{313}Ts → \textsuperscript{309}Mc & 8.31 & 0 & 6.62 & 6.12 & 7.47 & 6.74 & \textsuperscript{311}119 → \textsuperscript{307}Ts & 10.60 & 0 & 3.48 & 0.68 & 4.06 & 3.32 \\
\textsuperscript{315}Ts → \textsuperscript{311}Mc & 8.91 & 5 & 5.84 & 3.88 & 5.30 & 4.49 & \textsuperscript{313}119 → \textsuperscript{309}Ts & 9.50 & 2 & 4.28 & 3.27 & 4.67 & 3.90 \\
\textsuperscript{317}Ts → \textsuperscript{313}Mc & 8.49 & 5 & 7.32 & 5.37 & 6.80 & 5.98 & \textsuperscript{315}119 → \textsuperscript{311}Ts & 9.31 & 1 & 4.79 & 3.43 & 4.87 & 4.07 \\
\textsuperscript{281}Qg → \textsuperscript{277}Lv & 13.93 & 5 & -5.60 & -7.74 & -6.56 & -6.73 & \textsuperscript{317}119 → \textsuperscript{313}Ts & 9.25 & 3 & 8.45 & 6.13 & 7.55 & 6.77 \\
\textsuperscript{282}Qg → \textsuperscript{278}Lv & 13.75 & 0 & -6.69 & -7.43 & -6.98 & -7.18 & \textsuperscript{319}119 → \textsuperscript{315}Ts & 8.49 & 6 & -5.68 & -7.28 & -6.03 & -6.23 \\
\textsuperscript{283}Qg → \textsuperscript{279}Lv & 13.59 & 7 & -4.44 & -7.15 & -5.95 & -6.15 & \textsuperscript{287}120 → \textsuperscript{283}Og & 13.92 & 4 & -6.22 & -6.98 & -6.47 & -6.71 \\
\textsuperscript{284}Qg → \textsuperscript{280}Lv & 13.29 & 0 & -5.91 & -6.60 & -6.13 & -6.35 & \textsuperscript{288}120 → \textsuperscript{284}Og & 13.75 & 0 & -5.94 & -6.87 & -5.59 & -5.82 \\
\textsuperscript{285}Qg → \textsuperscript{281}Lv & 13.07 & 0 & -5.33 & -6.18 & -4.96 & -5.18 & \textsuperscript{289}120 → \textsuperscript{285}Og & 13.68 & 0 & -6.03 & -6.79 & -6.24 & -6.52 \\
\textsuperscript{286}Qg → \textsuperscript{282}Lv & 12.89 & 0 & -5.20 & -5.84 & -5.35 & -5.60 & \textsuperscript{290}120 → \textsuperscript{286}Og & 13.63 & 0 & -5.30 & -6.39 & -5.08 & -5.34 \\
\textsuperscript{287}Qg → \textsuperscript{283}Lv & 12.73 & 0 & -4.71 & -5.53 & -4.29 & -4.53 & \textsuperscript{291}120 → \textsuperscript{287}Og & 13.41 & 2 & -5.47 & -6.21 & -5.64 & -5.94 \\
\textsuperscript{288}Qg → \textsuperscript{284}Lv & 12.52 & 0 & -4.51 & -5.11 & -4.60 & -4.87 & \textsuperscript{292}120 → \textsuperscript{288}Og & 13.31 & 0 & -4.97 & -6.09 & -4.75 & -5.04 \\
\textsuperscript{289}Qg → \textsuperscript{285}Lv & 12.44 & 0 & -4.18 & -4.96 & -3.69 & -3.96 & \textsuperscript{293}120 → \textsuperscript{289}Og & 13.24 & 2 & -5.10 & -5.76 & -5.17 & -5.50 \\
\textsuperscript{290}Qg → \textsuperscript{286}Lv & 12.41 & 0 & -4.26 & -4.91 & -4.37 & -4.67 & \textsuperscript{294}120 → \textsuperscript{290}Og & 13.07 & 0 & -4.74 & -5.84 & -4.48 & -4.80 \\
\textsuperscript{291}Qg → \textsuperscript{287}Lv & 12.22 & 2 & -3.53 & -4.51 & -3.22 & -3.52 & \textsuperscript{295}120 → \textsuperscript{291}Og & 13.10 & 2 & -5.27 & -6.04 & -5.40 & -5.78 \\
\textsuperscript{292}Qg → \textsuperscript{288}Lv & 12.01 & 0 & -3.54 & -4.06 & -3.51 & -3.83 & \textsuperscript{296}120 → \textsuperscript{292}Og & 13.19 & 0 & -4.82 & -5.71 & -4.32 & -4.67 \\
\textsuperscript{293}Qg → \textsuperscript{289}Lv & 12.02 & 2 & -3.14 & -4.10 & -2.78 & -3.11 & \textsuperscript{297}120 → \textsuperscript{293}Og & 13.02 & 0 & -4.74 & -5.49 & -4.83 & -5.23 \\
\textsuperscript{294}Qg → \textsuperscript{290}Lv & 11.97 & 0 & -3.37 & -4.01 & -3.42 & -3.78 & \textsuperscript{298}120 → \textsuperscript{294}Og & 12.90 & 0 & -5.16 & -6.09 & -4.66 & -5.05 \\
\textsuperscript{295}Qg → \textsuperscript{291}Lv & 11.70 & 0 & -2.67 & -3.40 & -2.06 & -2.41 & \textsuperscript{299}120 → \textsuperscript{295}Og & 13.19 & 0 & -5.58 & -6.30 & -5.60 & -6.05 \\
\textsuperscript{296}Qg → \textsuperscript{292}Lv & 11.56 & 0 & -2.50 & -3.09 & -2.48 & -2.86 & \textsuperscript{300}120 → \textsuperscript{296}Og & 13.29 & 0 & -4.69 & -5.82 & -4.36 & -4.78 \\
\textsuperscript{297}Qg → \textsuperscript{293}Lv & 12.00 & 0 & -3.34 & -4.13 & -2.74 & -3.13 & \textsuperscript{301}120 → \textsuperscript{297}Og & 13.04 & 2 & -4.85 & -5.51 & -4.78 & -5.26 \\
\textsuperscript{298}Qg → \textsuperscript{294}Lv & 12.12 & 0 & -3.83 & -4.41 & -3.75 & -4.19 & \textsuperscript{302}120 → \textsuperscript{298}Og & 12.88 & 0 & -4.25 & -5.36 & -3.87 & -4.32 \\
\textsuperscript{299}Qg → \textsuperscript{295}Lv & 11.99 & 2 & -3.13 & -4.14 & -2.71 & -3.15 & \textsuperscript{303}120 → \textsuperscript{299}Og & 12.80 & 2 & -4.63 & -5.28 & -4.52 & -5.03 \\
\textsuperscript{300}Qg → \textsuperscript{296}Lv & 11.91 & 0 & -3.43 & -3.97 & -3.28 & -3.75 & \textsuperscript{304}120 → \textsuperscript{300}Og & 12.75 & 0 & -5.17 & -6.35 & -4.81 & -5.31 \\
\textsuperscript{301}Qg → \textsuperscript{297}Lv & 11.98 & 2 & -3.14 & -4.15 & -2.69 & -3.16 & \textsuperscript{305}120 → \textsuperscript{301}Og & 13.27 & 2 & -6.64 & -7.42 & -6.59 & -7.17 \\
\textsuperscript{302}Qg → \textsuperscript{298}Lv & 12.00 & 0 & -3.66 & -4.21 & -3.48 & -3.99 & \textsuperscript{306}120 → \textsuperscript{302}Og & 13.82 & 0 & -5.30 & -6.98 & -5.40 & -5.94 \\
\textsuperscript{303}Qg → \textsuperscript{299}Lv & 12.55 & 4 & -3.86 & -5.43 & -3.91 & -4.43 & \textsuperscript{307}120 → \textsuperscript{303}Og & 13.58 & 4 & -5.26 & -5.94 & -5.10 & -5.70 \\
\textsuperscript{304}Qg → \textsuperscript{300}Lv & 13.10 & 0 & -5.88 & -6.56 & -5.76 & -6.35 & \textsuperscript{308}120 → \textsuperscript{304}Og & 13.04 & 0 & -3.28 & -4.10 & -2.53 & -3.07 \\
\textsuperscript{305}Qg → \textsuperscript{301}Lv & 12.93 & 2 & -5.10 & -6.24 & -4.68 & -5.25 & \textsuperscript{309}120 → \textsuperscript{305}Og & 12.17 & 0 & -2.01 & -2.50 & -1.69 & -2.27 \\
\textsuperscript{306}Qg → \textsuperscript{302}Lv & 12.53 & 0 & -4.82 & -5.43 & -4.62 & -5.22 & \textsuperscript{310}120 → \textsuperscript{306}Og & 11.48 & 0 & -0.61 & -1.56 & 0.00 & -0.53 \\
\textsuperscript{307}Qg → \textsuperscript{303}Lv & 11.99 & 2 & -3.24 & -4.27 & -2.71 & -3.28 & \textsuperscript{311}120 → \textsuperscript{307}Og & 11.10 & 2 & -1.00 & -1.45 & -0.62 & -1.22 \\
\textsuperscript{308}Qg → \textsuperscript{304}Lv & 11.20 & 0 & -1.96 & -2.41 & -1.61 & -2.20 & \textsuperscript{312}120 → \textsuperscript{308}Og & 11.05 & 0 & -0.17 & -1.12 & 0.46 & -0.10 \\
\textsuperscript{309}Qg → \textsuperscript{305}Lv & 10.67 & 2 & -0.13 & -1.05 & 0.49 & -0.07 & \textsuperscript{313}120 → \textsuperscript{309}Og & 10.92 & 2 & 1.90 & 1.55 & 2.35 & 1.77 \\
\textsuperscript{310}Qg → \textsuperscript{306}Lv & 10.29 & 0 & 0.35 & -0.02 & 0.77 & 0.18 & \textsuperscript{314}120 → \textsuperscript{310}Og & 9.97 & 0 & 5.05 & 4.18 & 5.70 & 5.19 \\
\textsuperscript{311}Qg → \textsuperscript{307}Lv & 9.39 & 2 & 3.56 & 2.70 & 4.21 & 3.67 & \textsuperscript{315}120 → \textsuperscript{311}Og & 10.15 & 0 & 2.18 & 1.51 & 2.35 & 1.73 \\
\textsuperscript{312}Qg → \textsuperscript{308}Lv & 9.06 & 0 & 4.07 & 3.79 & 4.55 & 3.98 & \textsuperscript{316}120 → \textsuperscript{312}Og & 9.97 & 0 & 2.87 & 1.53 & 3.13 & 2.55 \\
\textsuperscript{313}Qg → \textsuperscript{309}Lv & 8.76 & 0 & 5.69 & 4.83 & 6.33 & 5.80 & \textsuperscript{317}120 → \textsuperscript{313}Og & 9.96 & 3 & 2.10 & 2.29 & 1.64 & -1.32 \\
\textsuperscript{314}Qg → \textsuperscript{310}Lv & 8.50 & 0 & 6.38 & 5.77 & 6.53 & 5.96 & \textsuperscript{318}120 → \textsuperscript{314}Og & 9.99 & 0 & 1.95 & 2.281 & 4.839 & 5.149 \\ \hline
\end{tabular}
}
\label{tab:alpha_decay_parameters} 
\end{table*}

\begin{figure}[htp]
  \centering
  
  \begin{subfigure}{0.5\textwidth}
    \centering
    \includegraphics[width=\linewidth, height=0.3\textheight, keepaspectratio]{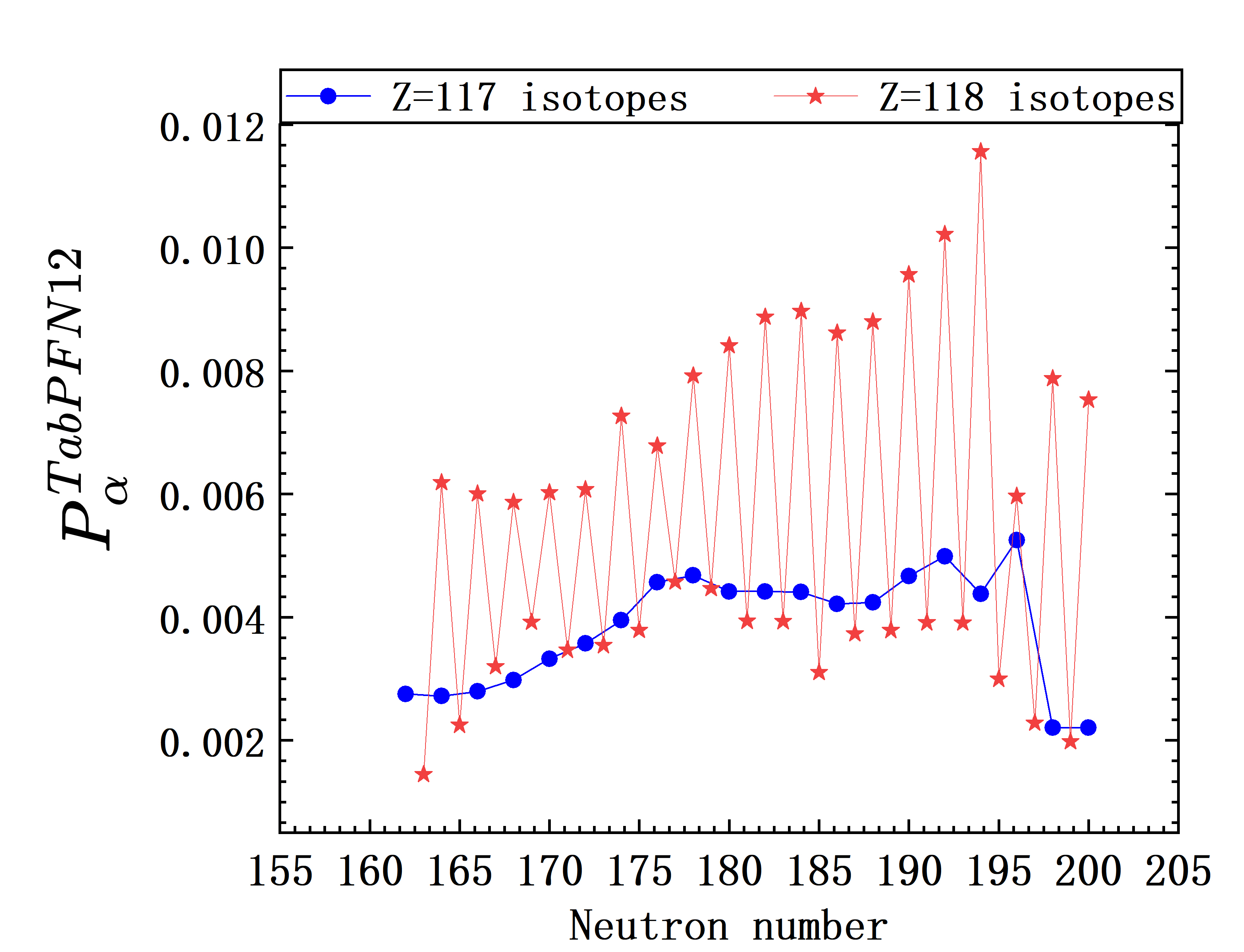}
  \end{subfigure}
  
  \vspace{10pt}  
  
  \begin{subfigure}{0.5\textwidth}
    \centering
    \includegraphics[width=\linewidth, height=0.3\textheight, keepaspectratio]{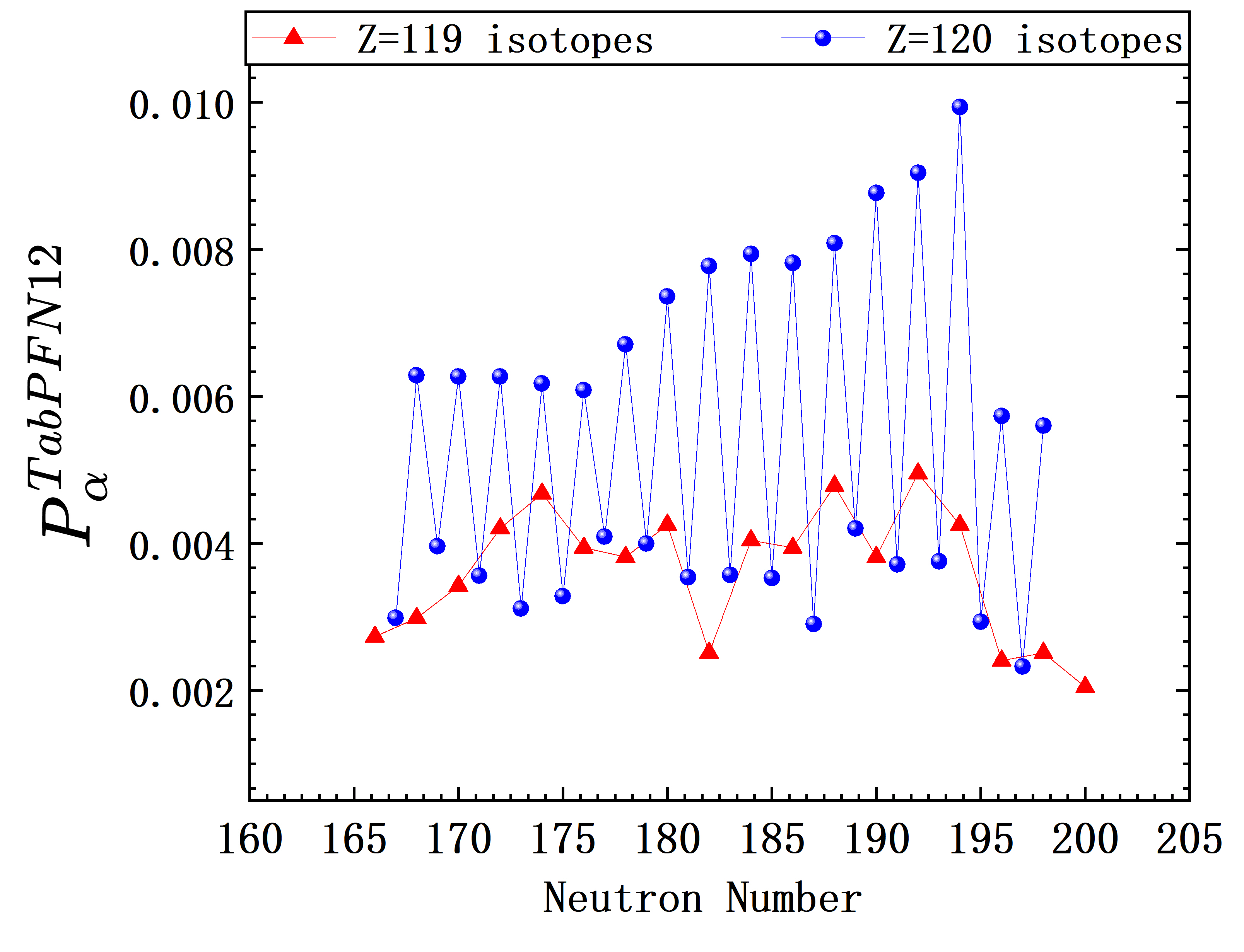}
  \end{subfigure}

  \caption{Upper panel: Predicted $\alpha$-particle preformation factors $P_{\alpha}^{\mathrm{TabPFN12}}$ for isotopic chains of $Z = 117$ and 118. Lower panel: Corresponding predictions for $Z = 119$ and 120 isotopic chains.}
  \label{fig:5}
\end{figure}

Accurate prediction of $\alpha$-decay half-lives is crucial for both the synthesis of superheavy elements and the understanding of their structural properties. The thoroughly trained TabPFN12 model demonstrates reliable predictive capability for $\alpha$-particle preformation factors. This model is therefore employed to predict the preformation factors for superheavy nuclei with $Z = 117$–120, utilizing $\alpha$-decay energies $Q_\alpha$ from Ref. \cite{ma2019basic}. Figure~\ref{fig:5} (a) presents the predicted preformation factors for $Z = 117$ and 118 isotopes, revealing a distinct odd-even staggering (OES) effect. For instance, at $N = 182$, $P_{\alpha}^{\mathrm{TabPFN12}}$ for $Z = 118$ is larger than for its $Z = 117$ isotone, decreasing from $\sim$0.0088 to 0.0044—a reduction of 50\%. Similarly, within the $Z = 118$ chain, $P_{\alpha}^{\mathrm{TabPFN12}}$ for $N = 182$ exceeds that for $N = 183$ (0.0039). A similar systematic behavior is observed for $Z = 119$ and 120 nuclei, as shown in Fig.~\ref{fig:5} (b). These results indicate that both unpaired protons and unpaired neutrons inhibit $\alpha$-particle preformation. Furthermore, the $P_{\alpha}^{\mathrm{TabPFN12}}$ values for the even-$Z$ chains (120, 118) are systematically larger than for the odd-$Z$ chains (119, 117), highlighting the key role of proton pairing correlations in $\alpha$-cluster preformation.

\begin{figure*}[!htbp]
    \centering
    \includegraphics[width=1\linewidth]{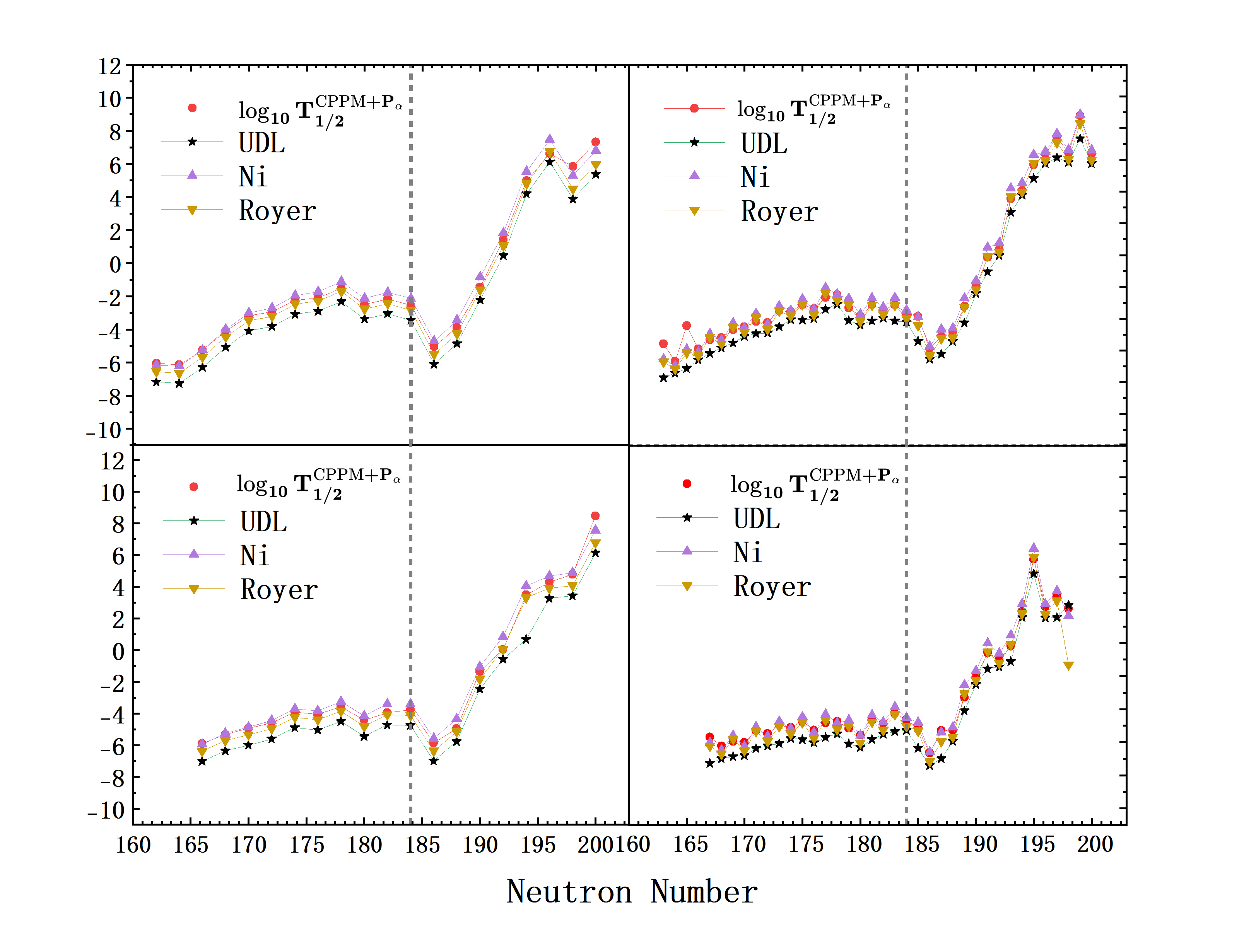}
    \caption{Predicted logarithmic $\alpha$-decay half-lives for nuclei with $Z = 117, 118, 119,$ and $120$, obtained using the TabPFN12 model and three empirical/semi-empirical approaches. The vertical dashed line indicates the neutron number $N=184$.}
    \label{fig:6}
\end{figure*}

The $P_{\alpha}^{\mathrm{TabPFN12}}$ values are incorporated into the CPPM to compute $\alpha$-decay half-lives in the superheavy region. Table IV summarizes the calculated logarithmic half-lives for $Z = 117$–120 isotopes using the $P_\alpha^{\mathrm{TabPFN12}} + \mathrm{CPPM}$ approach, alongside predictions from the Universal Decay Law (UDL), Ni's empirical formula, and the Royer semi-empirical formula. The first four columns list the decay process, $Q_\alpha$ value, orbital angular momentum $l$ of the $\alpha$ particle, and quadrupole deformation parameter $\beta_2$ of the parent nucleus. Columns 5–8 present the logarithmic half-lives predicted by the different models. Fig.~\ref{fig:6} compares the $P_\alpha^{\mathrm{TabPFN12}} + \mathrm{CPPM}$ results with predictions from the empirical and semi-empirical formulas. Good agreement is observed among all models for nuclei with $Z = 117$–120.

Notably, the $\alpha$-decay half-lives decrease sharply from $N = 184$ to $N = 186$, dropping by over two orders of magnitude with the addition of only two neutrons. This rapid change suggests that $N = 184$ may represent a new neutron magic number beyond $N = 126$. It is important to note that shell effects are implicitly captured by the TabPFN12 model from the data—similar to its treatment of $Q_\alpha$ and deformation—as shell correction terms are not explicitly included in the input features.

\section{summary}

This work develops a hybrid approach combining the Tabular Prior-data Fitted Network (TabPFN) with the Coulomb and Proximity Potential Model (CPPM) to systematically investigate $\alpha$-particle preformation factors and their impact on $\alpha$-decay half-life predictions. The TabPFN model, a transformer-based architecture pretrained on diverse synthetic datasets, learns the complex mapping between nuclear structure properties and $\alpha$-preformation probabilities directly from experimental data.

Systematic analysis demonstrates that incorporating key nuclear structure features—including pairing effects ($\delta$), proton and neutron parities ($Z_p$, $N_p$), angular momentum ($l$), and quadrupole deformation ($\beta_2$)—significantly enhances the predictive accuracy for $\alpha$-preformation factors. The comprehensive TabPFN12 model, incorporating all nine physical descriptors, achieves a root-mean-square deviation of $\sigma_{\mathrm{rms}} = 0.211$ against experimental preformation factors extracted from 498 nuclei, representing improvements of 63.8\% and 66.8\% over existing empirical formulas.

The predicted preformation factors exhibit clear dependencies on nuclear structure, including pronounced odd-even staggering and shell closures at $Z = 82$ and $N = 126$. The TabPFN12 model successfully captures the linear correlation between $\log_{10}P_\alpha$ and $Q_\alpha^{-1/2}$, effectively extending the Geiger-Nuttall systematics to preformation factors. A clear negative correlation between $\log_{10}P_{\alpha}$ and the fragmentation potential $V_{\mathrm{frag}}$ is also observed. The results confirm that unpaired nucleons significantly inhibit $\alpha$-cluster formation, as evidenced by systematically lower preformation factors in odd-$A$ nuclei compared to their even-even neighbors.

Incorporated into the CPPM framework, the TabPFN12-predicted preformation factors substantially improve $\alpha$-decay half-life calculations, reducing the RMS deviation from 2.028 to 0.211 (89.5\% improvement) for the training set and from 2.920 to 0.771 (73.5\% improvement) for 41 newly evaluated superheavy nuclei. A similar improvement is also obtained when a deformed Coulomb plus Woods–Saxon potential is adopted, yielding an rms deviation of $\sigma_{\mathrm{rms}} = 0.232$, which remains close to the value obtained within the CPPM framework and demonstrates the robustness of the TabPFN-derived preformation factors against different potential descriptions. The model's extrapolation capability is further validated by predictions for $Z = 117$--120 isotopic chains, which reveal a sharp decrease in half-lives from $N = 184$ to $N = 186$. This rapid change suggests that $N = 184$ may constitute a new neutron magic number in the superheavy region.

This study establishes the TabPFN framework as a powerful tool for extracting nuclear structure information from decay data and for providing reliable predictions to guide superheavy-nuclei synthesis experiments. The hybrid machine learning approach presented herein offers a promising pathway for refining nuclear models and for exploring uncharted regions of the nuclear landscape.
\\
\section{Acknowledgements}

This work is supported by Yunnan Provincial Science Foundation Project (No. 202501AT070067), Yunnan Provincial Xing Dian Talent Support Program (Young Talents Special Program, No. XDYC-QNRC-2023-0162), Kunming University Talent Introduction Research Project (No. YJL24019), Yunnan Provincial Department of Education Scientific Research Fund Project (No. 2025Y1042 and 2025Y1055), the Program for Frontier Research Team of Kunming University 2023, National Natural Science Foundation of China (No. 12063006), the Special Basic Cooperative Research Programs of Yunnan Provincial Undergraduate Universities’ Association (grant NO. 202101BA070001-144), and Xing Dingyu Academician Workstation of Yunnan Province (No. 202605AF350035).

\nocite{*}

\bibliography{apssamp}

\end{document}